\documentclass[manuscript]{aastex}

\shorttitle{First Results from RAO Variable Star Program}
\shortauthors{Williams and Milone}

\begin{document}

\title{First Results from the RAO Variable Star Search Program. \\ 
I. Background, Procedure, and Results from RAO Field 1\footnotemark[1]}

\author{Michael D. Williams}
\affil{Physics and Astronomy Department, University of Calgary}
\email{williamd@alumni.ucalgary.ca}  

\and

\author{E. F. Milone}
\affil{Physics and Astronomy Department, University of Calgary}
\email{milone@ucalgary.ca}
\footnotetext[1]{Publications of the RAO No. 76}
\begin{abstract}
We describe an ongoing variable star search program and present the first
reduced results of a search in a 19 square degree (4.4$\arcdeg$~$\times$~4.4$\arcdeg$) 
field centered on J2000 $\alpha$ = 22:03:24, $\delta$ = +18:54:32.  
The search was carried out with the Baker-Nunn Patrol Camera located at the 
Rothney Astrophysical Observatory in the foothills of the Canadian Rockies.  
A total of 26,271 stars were detected in the field, over a range of about 11-15 
(instrumental) magnitudes. 
Our image processing made use of the IRAF version of the DAOPHOT aperture 
photometry routine and we used the ANOVA method to search for periodic variations 
in the light curves.  
We formally detected periodic variability in 35 stars, that we tentatively
classify according to light curve characteristics: 6 EA 
(Algol), 5 EB ($\beta$ Lyrae), 19 EW (W UMa), and 5 RR (RR Lyrae) stars.
Eleven of the detected variable stars have been reported previously in the literature.  
The eclipsing binary light curves have been analyzed with a package of light curve
modeling programs and 25 have yielded converged solutions. Nine of these systems are
detached, 3 semi--detached, 10 over--contact, and 2 appear to be in marginal contact.
We discuss these results as well as the advantages and disadvantages of the 
instrument and of the program.
\end{abstract}


\section{Introduction}
We have used a former Baker-Nunn satellite tracking camera, refurbished as the 
{\it Baker-Nunn Patrol Camera} ({\it BNPC}), at the {\it Rothney Astrophysical 
Observatory} ({\it RAO}) of the University of Calgary, to acquire data over 
several years in a search for photometrically variable objects.
In this first paper we describe the search program, the challenges that have
had to be met to obtain useful data, and present the results for a field 
centered on the J2000 coordinates $\alpha$ = 22:03:24, $\delta$  = +18:54:32, 
that we observed over the interval September 23 to October 27, 2004.
 
The BNPC is an f/0.96 camera with a 60 cm primary mirror diameter and a 50 cm 
diameter corrector plate.  The refurbishment was carried out mainly by FMB of 
Boulder, Colorado, with a good deal of local design and engineering work.   
The instrument is now mounted equatorially, without
the third axis that permitted the tracking of non-equatorial Earth satellite orbits.  

\begin{figure}
\plotone{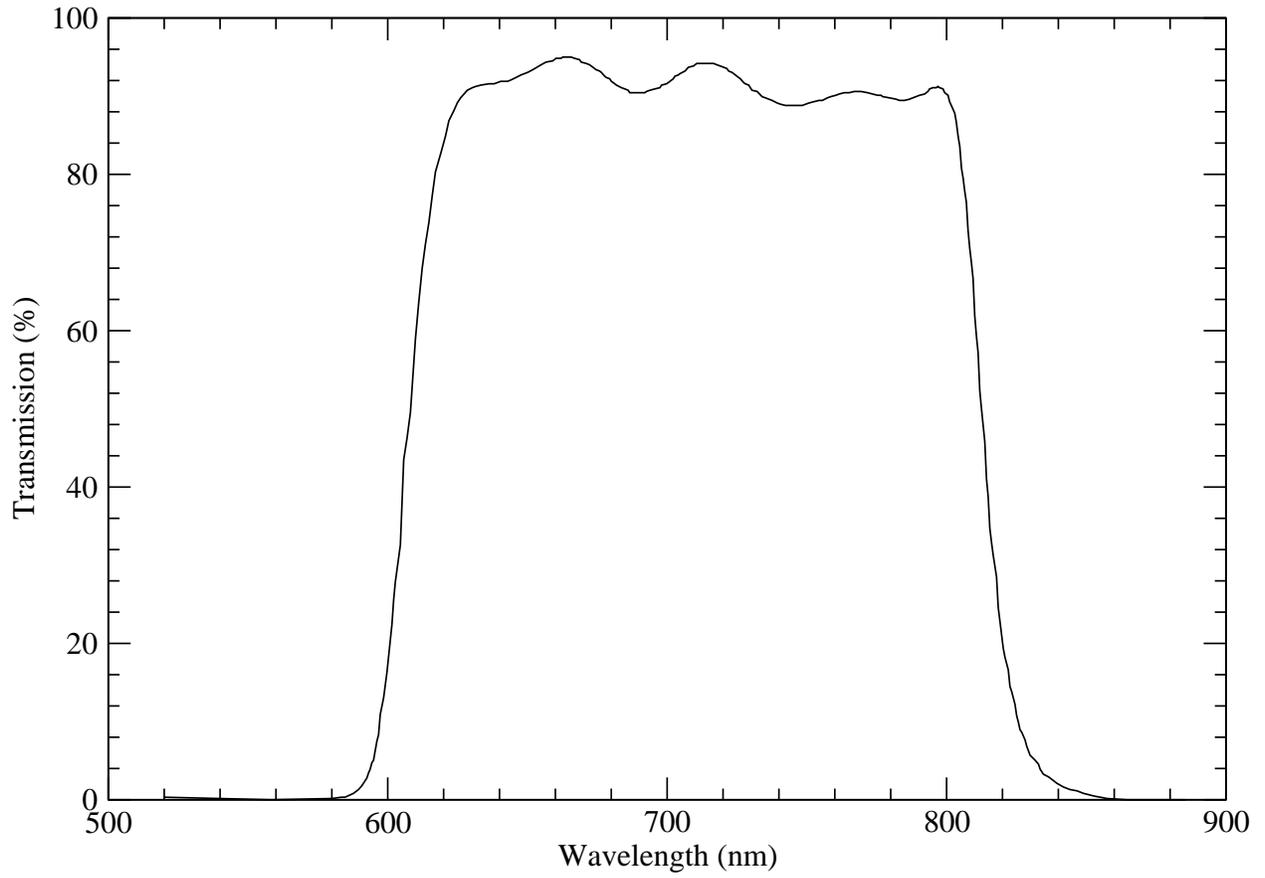}
\caption{The transmission curve for the filter used on the BNPC.\label{filtertran}}
\end{figure}

The detector is a 4096 x 4096 front-illuminated, KAF-16801E chip with
transparent gate (and thus a relatively high sensitivity for such a chip)
and pixel dimensions of 9 $\times$ 9 $\mu$m, in an FLI camera.  
The image scale is 3.89$\arcsec$/pixel for a field of 
view of 19.6 square-degrees.  The camera is equipped with a single filter,
with peak transmission $\sim705$ nm and HPFW $\sim205$ nm, properties similar
to those of the Johnson $R$ passband [700 and 200 nm, respectively;
Drilling \& Landolt (2000)\nocite{drillinglandolt00}].  See Figure~\ref{filtertran} 
for a plot of the filter transmission, measured and replotted from a manufacturer 
supplied tracing.  

\section{The Variable Star Search Program}
Previous survey work has involved searches for transient objects, and some 
have specifically targeted gravitational lensing events such as {\it OGLE} 
[FOR {\it Optical Gravitational Lens Experiment}; Udalski et al., (1993, 2002)]\nocite{Udal93}\nocite{Udal02};
and {\it MACHO} [for {\it Massive Cold Halo Objects}; Alcock et al. (1995)\nocite{Alcock95}]. 
Others
have targeted extrasolar planet transits, such as {\it HATNet} [for \textit{Hungary Automatic 
Telescope Network}; Bakos et al. (2002, 2004a)]\nocite{Bakos02}\nocite{Bakos04} 
and {\it TrES} [for \textit{Transatlantic Extrasolar Planet Survey}; Alfonso (2004)\nocite{Alonso04}], 
as well as variable stars.  One of the more comparable variable star surveys is the 
{\it MOTESS-GNAT}
[for {\it Moving Object and Transient Event Search System and Global Network of
Astronomical Telescopes}] survey (Kraus et al. 2007)\nocite{Kraus07}, although it 
currently involves three telescopes of 35-cm aperture, instead of
one 50-cm telescope, as for the BNPC. 
Most such surveys make use of either small resolution, large field cameras
with relatively bright faintness limits, as HATNet, or larger telescopes with smaller--field 
image cameras, as MACHO.

The BNPC has the great advantage of enabling a search field of more than 19 
square degrees of sky per exposure, and yet of reaching relatively faint limits,
so that it provides excellent opportunities to sample the variable star 
populations across large swaths of the sky. 

As Kraus et al. (2007)\nocite{Kraus07} note, programs such as this fill 
``... a unique niche in parameter space, observing with intermediate depth, ...'' 
as well as large area and flexible repetition strategy.

The initial field selected for the program contains two open clusters but its 
selection had more to do with its serendipitous placement in the sky when the 
BNPC had been sufficiently well adjusted to permit commencement of observations.
This field is well off the galactic plane and more or less
devoid of very bright stars, although not devoid of objects of interest.  It
contains, for example, the planetary transiting system HD~209458, regrettably
overexposed in the images images reported on here, as are most of the known 
variable stars in this field.  

Other fields we are observing include the extensive
Hyades and Coma star cluster regions and selected regions at various galactic 
latitudes.  The regions away from the plane can be expected to produce 
an increased relative number of population II variables, and, because of the 
inclusion of external galaxies, supernovae.  The frames are being archived for
future investigations.  

Our main goal in this program,
however, is to improve the database of fundamental parameters of stars
through investigation of eclipsing and regularly pulsating variable stars.
Table~\ref{all_plates} summarizes the data obtained to date which are 
in various stages of image reduction and data analysis.

Limitations on variable star detections are discussed below, with emphasis on 
those found in the analysis of the first set of data of what we call \textit{RAO 
Field 1}. These and subseqent data sets will be made available on--line.

\section{Observations and Observational Constraints}

The observations reported here were obtained over the JDN interval 2453271.5 
to 2453305.5.  The summary of these data is presented in Table~\ref{observationslist}.  
The differences in numbers of frames in Tables 1 and 2 are due to relatively poor
focus for some of the frames counted in Table 1 which were omitted from the analysis
discussed here.

By trial and error we find that in order to detect variations of a few 
percent or less, the effective faint raw magnitude limit is $R_J \approx{17}$,
determined by the sky background and concomitant noise.  
The bright end of the range, $\sim{11.3}$, is established by saturation limits.  
The highest precision is to be expected for the brighter side of this range, and it is 
there where an extrasolar planetary transit may be detected, although this 
still relatively rare type of detection event is not the primary focus of our program.  

The skies over the RAO, located in the foothills of the Canadian Rockies at 
an elevation of 1300 m, are frequently non-photometric in the 
classical photoelectric photometry sense.  The observatory is located within 50 km of 
the outskirts of a major metropolitan city with large if increasingly ameliorated 
light pollution (the introduction of lower wattage, downward-projecting street lamps
being the ameliorating factor).  For this reason, the Rapid
Alternate Detection System ({\it RADS}) for gated, pulse-counting differential
photoelectric photometry of individual variable stars, was developed here 
(Milone \& Robb 1983)\nocite{Mil83} because it eliminates first order extinction variation 
as well as sky brightness and its variation, except that occurring at the chopping 
frequency of the secondary mirror.  
One might expect that CCD photometry would allow one to correct for some of these 
effects, and, for this reason, comparison stars were taken as close to each variable
candidate as was practical, as we describe below. However, the large field of view 
and small image scale have so far conspired to produce uneven results.  
Consequently, the nights we can actually use tend to be photometric in the classical
sense of the word.  Sky conditions, compounded by scheduling gaps, therefore, limit 
the precision of the data we have been able to acquire.

When the sky quality is high enough to justify observing, the large area 
detector CCD provides data on a large number of stars.
The downside of the large sky field and consequent low spatial resolution is
that even relatively sparse fields have to be treated as crowded ones.
Thus, instead of the much-preferred multi-aperture photometry that involves 
what Stetson (1987)\nocite{Ste87} has called a ``curve-of-growth'' procedure, we have been 
able to apply only single aperture photometry.  PSF fitting is now seen to be 
needed in order to reduce the effects of neighboring stars.  However, PSF variations 
across the field produce additional complications, as we discuss below.

\section{Image Analysis and Data Reduction}

\subsection{Image and Data Processing}

The DAOPHOT package (Stetson 1987, 1990) within the IRAF data reduction package
was used for the image processing.  MDW wrote scripts to facilitate image processing 
and reduction calculations and to organize the results.
The magnitudes were determined with a single 9-pixel aperture.  

The raw magnitude of each star was converted into a differential magnitude with 
respect to a mean of $\sim{10}$ candidate comparison stars in the vicinity of 
each star image.  
The comparison star selection was automated and each of the ten had to meet the 
a number of criteria designed to assure constancy within precisional limits.  
Each candidate star needed to:

\begin{itemize}
\item have a signal-to-noise ratio better than the average;  
\item display differential light curves (with respect to the other stars) 
that deviate insignificantly from constant light;  
\item lay within 500 pixels of the program star; and
\item produce differential light curves with the lowest standard deviation. 
\end{itemize}

The last criterion, especially, has limited the number of comparison stars.

The differential magnitude was calculated from Equation~\ref{diffmagequ}.

\begin{equation}
\label{diffmagequ}
DM_{i} = \frac{\sum_{j=0}^{j<n}{(M_{i}-k'X)-(M_{i,j}^{c}-k'_{i,j}X_{i,j}^{c})}}{n}
\end{equation}

where $i$ indexes the candidate variable star, $j$ indexes comparison star, 
$n$ is the number of comparison stars (10, if possible, to beat down the effects
of low amplitude stellar variation and such observational effects as residual psf 
variations across the field), $X$ is airmass, 
$k'$ is linear extinction coefficient, and $M$, and $DM$ are the magnitude 
and the differential magnitude, respectively; the superscript $c$ refers to 
comparison stars.

\subsection{Data Reduction}
Our data have been corrected for extinction and have been approximately 
transformed to Johnson $R$ magnitudes.

For each night we calculated the Bouguer\nocite{Bou1729} extinction coefficient 
(Bouguer 1729)\nocite{Bou1729}, $k\arcmin$, 
for each comparison star with a linear least squares fit and then performed a 
sigma clipped average of all the individual star $k\arcmin$ values to find 
the nightly $k\arcmin$ mean.

The zero point of the magnitude scale was determined by comparing our systemic 
magnitudes and the $R$ magnitudes tabulated in the USNO 2.0A catalog.  There
seemed to be no sensitivity to color in this zero point correction, as, for example,
the color index term expressions given by Hardie (1962)\nocite{Hardie62}.  

\begin{figure}
\plottwo{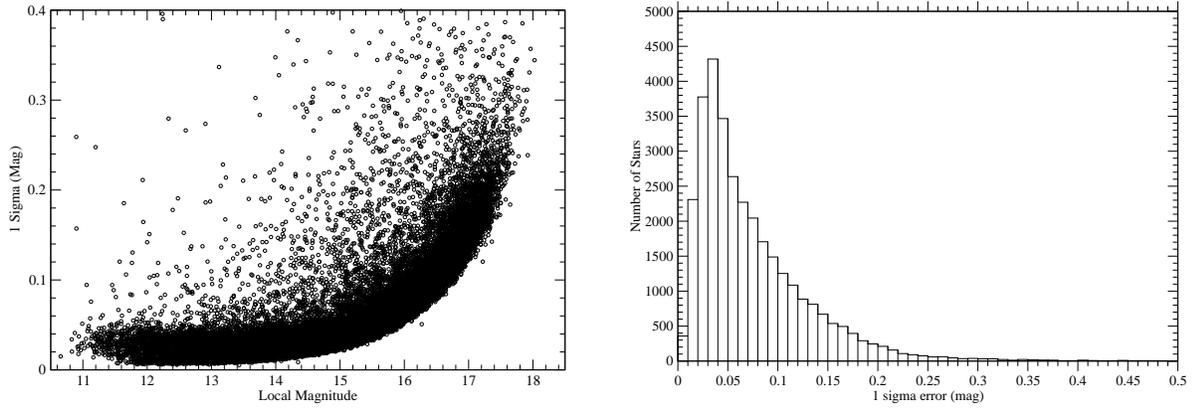}{williams_milone_Fig02b.eps}
\caption{The precision as a function of magnitudes for the measured light 
curves (left) and the histogram of achieved 1 $\sigma$ levels in the light 
curves (right)\label{hist}}
\end{figure}

Figure~\ref{hist} is a histogram of the 1$\sigma$ levels in the light curves as
determined thus far.

\section{Variability Detection and Analysis}\label{secdetect}

The differential light curves were searched for periodic variations using the 
analysis of variations ({\it ANOVA}) technique outlined in Davies (1990, 1991)
\nocite{Davies90}\nocite{Davies91}.  This technique searches for periods with which a 
significant variation in mean values between different phase bins is produced.

Before each light curve was searched, it was clipped of points that exceeded
5$\sigma$ from the mean.  If more than four points exceeded this limit, no points 
were clipped, limiting the clipping to solely erratic points.

\begin{figure}
\plottwo{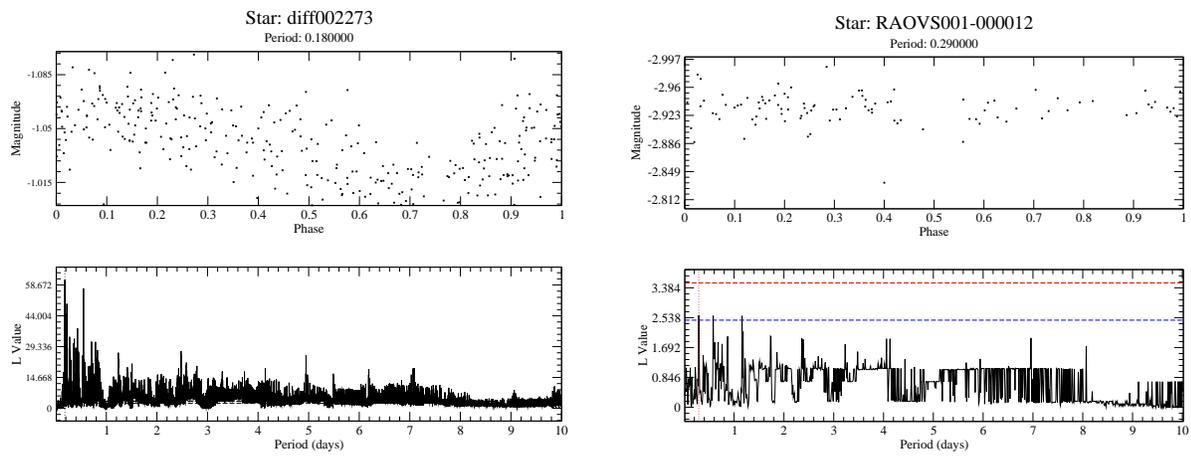}{williams_milone_Fig03b.eps}
\caption{Two light curves and their L statistic plots that illustrate
cases where variability is detected (left) and where no variability is detected 
(right).  The phased light curves are plotted in the top graph.  
The horizontal dashed lines in the L statistic plots represent the 99\% and 
95\% confidence levels for the lower (blue) and upper (red) lines, respectively
(colors visible in the on-line version of this paper).\label{lvalues}}
\end{figure}

The search produces an L statistic value (Davies 1990)\nocite{Davies90} for each period 
searched (for examples see Figure~\ref{lvalues}); the larger the L statistic 
the greater the variability between phase bins.  The test statistic follows an 
F distribution, which is used to determine if the L statistic value is indeed 
statistically significant.  If the maximum test statistic for a certain light 
curve is significant then the detection routine marks that star as variable and 
records the period associated with the maximum as the period of variation. 

The light curves were searched for a range of periods between 0.01 days and 10 
days, in increments of 0.001 day.  The phased light curves were divided into 5 
bins and we looked for variations in the means among the bins to a 
significance level of 99\%.  For RAO Field 1, we initially detected apparent 
variability in 10,406 of the 26,271 stars searched.  Such a high fraction of 
apparent variability was unexpected and disappointing.  Thus the stars that were 
detected as variable had to be sifted further, individually.  Most of this
variability was found to be due to systematic noise sources. 
In many cases the systematic noise was due to light contamination from 
nearby stars and this can vary from frame to frame, primarily due to seeing 
variations. 

The FWHM of a stellar profile can vary by as much as a 
factor 2 during a night.  As the stellar FWHM increases, the level of 
contamination seen in the fixed aperture also increases: apparent
variability in the light curves of stars that have bright near neighbors is 
induced by variability in their stellar profiles.  

Typical FWHMs are 2.34 pixels, equivalent to 9.10$\arcsec$, at the center of 
the frame, and 3.77 pixels, or 14.67$\arcsec$, at the corners, but they do not 
vary smoothly across the image, and, in fact, the spatial variation differs 
during a night's observing.  Seeing (typically 3--4$\arcsec$ but variable) 
contributes to the large star images, but there appears to be an important
instrumental contribution also.

To date, PSF fitting has proven both difficult and time consuming to apply 
reliably, due to the PSF variability.  Nevertheless, the large number of stars 
that display systematic noise in light curves requires continued use of this 
method to reduce the effects of neighboring stars.

\begin{figure}
\plotone{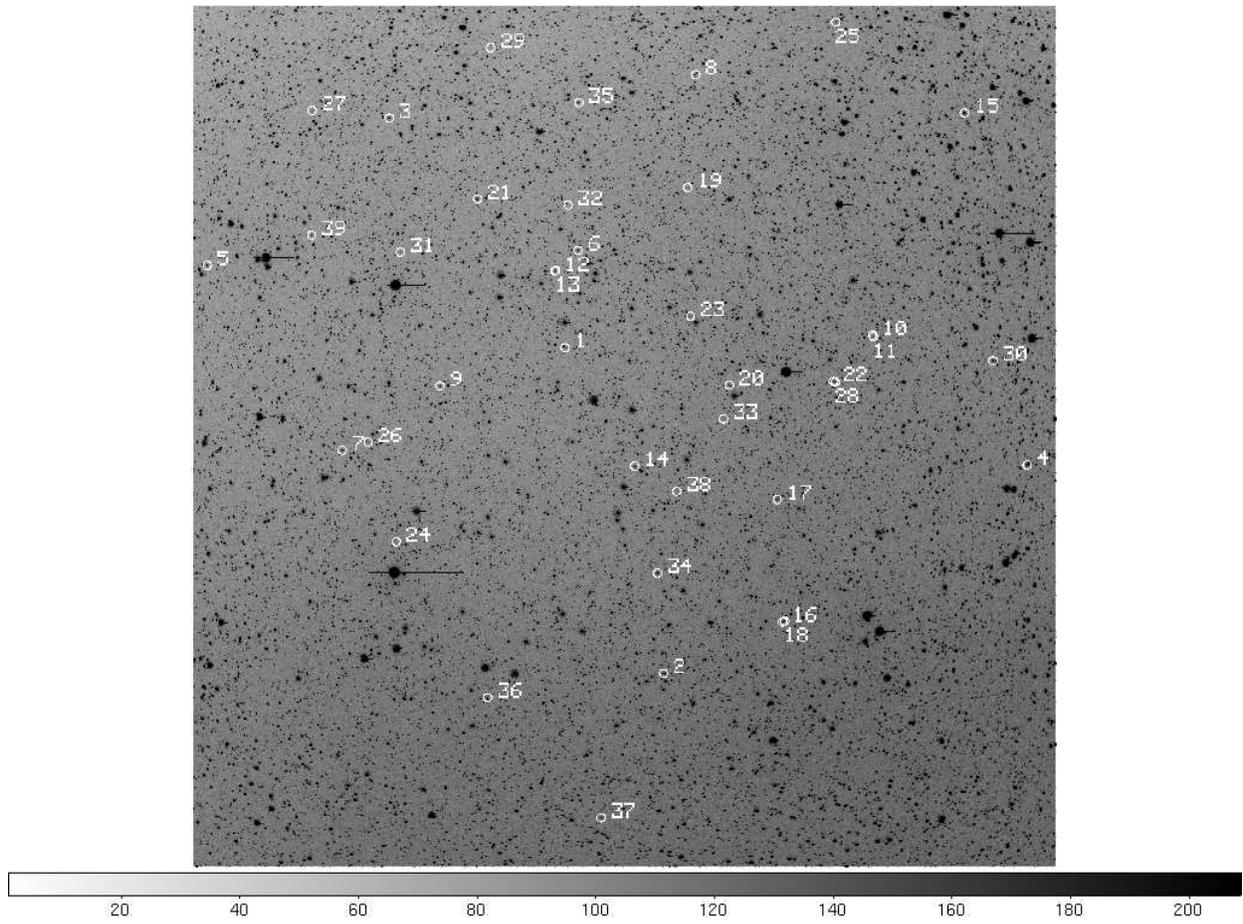}
\caption{RAO Field 1 showing the detected variable stars \label{fieldphoto}}
\end{figure}

The detected variables and their determined properties are listed in 
Table~\ref{starlist}.  They are marked and labeled in Figure \ref{fieldphoto}.
The ``Auto Period'' is the period detected by the 
automatic search routine.  The uncertainty (mse) is given in parentheses in 
units of the last decimal place.

The light curves of the detected variables were subsequently searched again, 
with more phase bins (40) and a finer period step (0.00001 day).  The peaks in 
the array of test statistics were fit with Gaussian curves to estimate 
periods; in each case the uncertainty was found from the HWHM of the peak. The period 
column in Table~\ref{starlist} records this refined period.  We obtained the epoch 
by fitting the deepest minimum (for binary stars) or the maximum (for 
pulsating variables) with a Gaussian function.  The uncertainty in the epoch 
was the formal error in this fitting.

However, the small image scale requires that we place caveats on the 
discoveries of several pairs of close--neighbor stars in the variable star list.
These include pairs RAO1-10 \& -11, -12 \& -13, -16 \& -18, and -22 \& -28.
With the exception of pair RAO1-16 \& -18, the stars making up these pairs had the same 
detection period, within the uncertainties.  It is thus likely that only one star in 
each such pair is the true variable.  The light from the variable star may contaminate 
the aperture of its non-variable neighbor of the pair, imparting its variation to an
otherwise non-variable light curve; and the variable, in turn, may be contaminated 
with light of the non--variable.  Both light curves will appear shallower then
the undiluted light curve because of the added light, in a flux-weighted sort of way.  
Assuming that each of the pairs of objects we identify as producing a common 
light curve does represent only one intrinsically variable star, we are left with 35 
apparently bonafide variable star detections.

\section{Results: The Variables in RAO Field 1}

\subsection{Periodic Variables Found and Not Found}\label{subsectypes}

\begin{figure}
\plotone{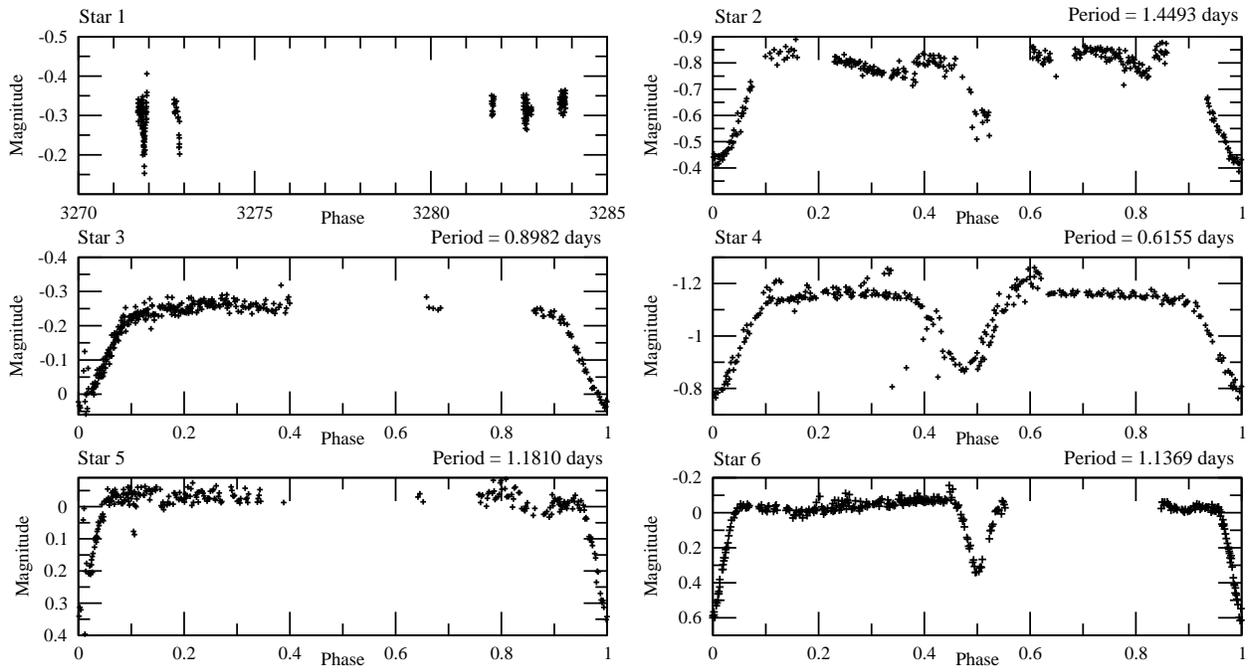}
\caption{Detected variable stars with Algol (EA)--like light curves.\label{algolfig}}
\end{figure}

\begin{figure}
\plotone{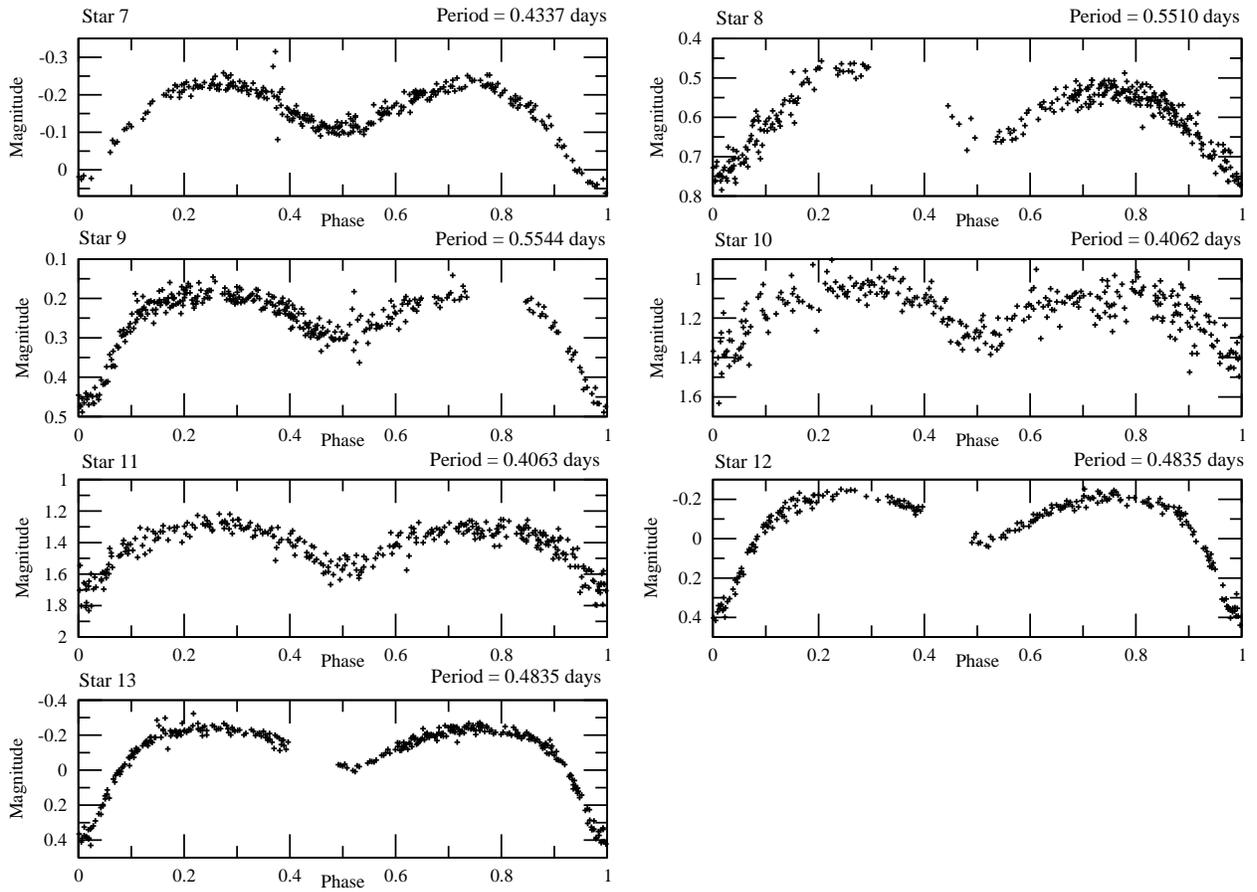}
\caption{Detected variables with $\beta$ Lyrae (EB) light curves.\label{blyrfig}}
\end{figure}

\begin{figure}
\plotone{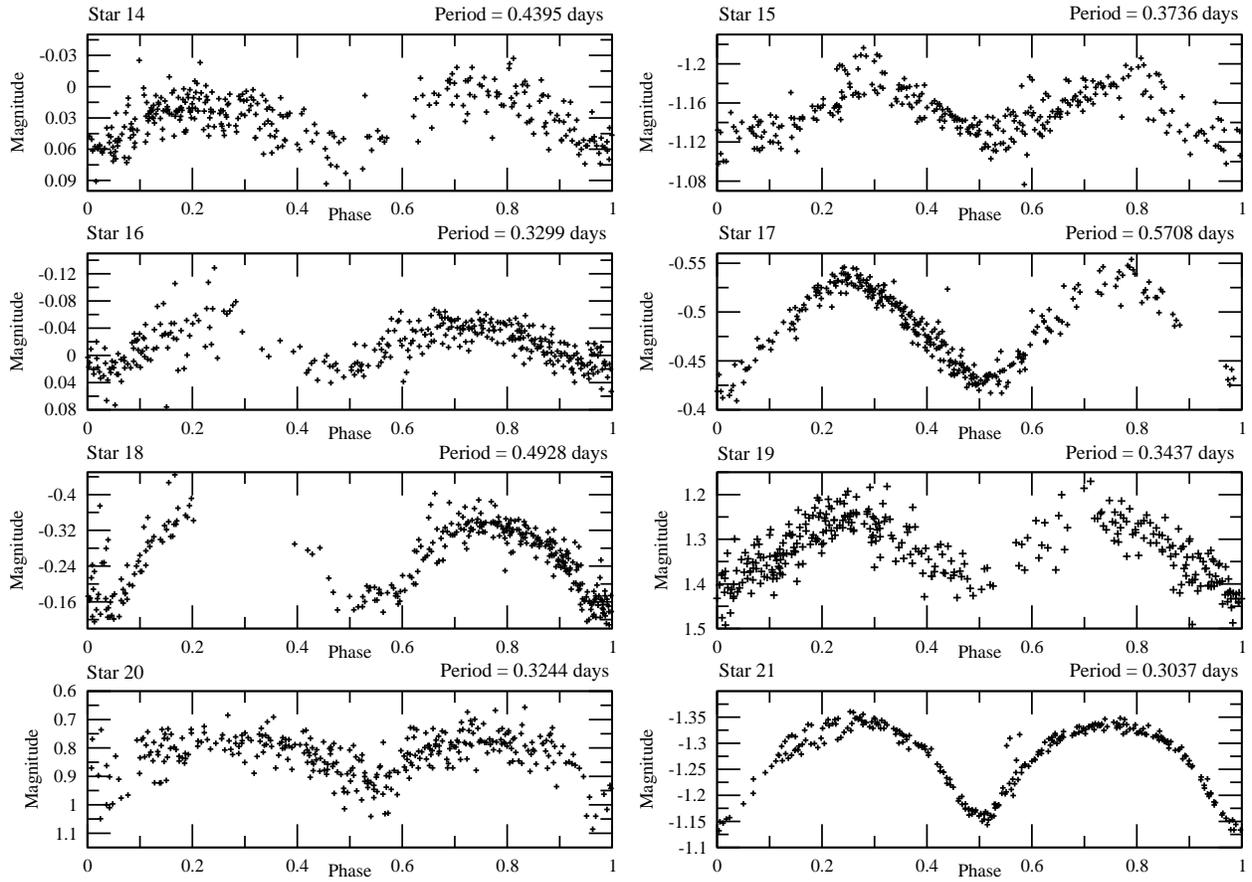}
\caption{Detected variable stars with W UMa (EW)--like light curves.\label{wuma1fig}}
\end{figure}

\begin{figure}
\plotone{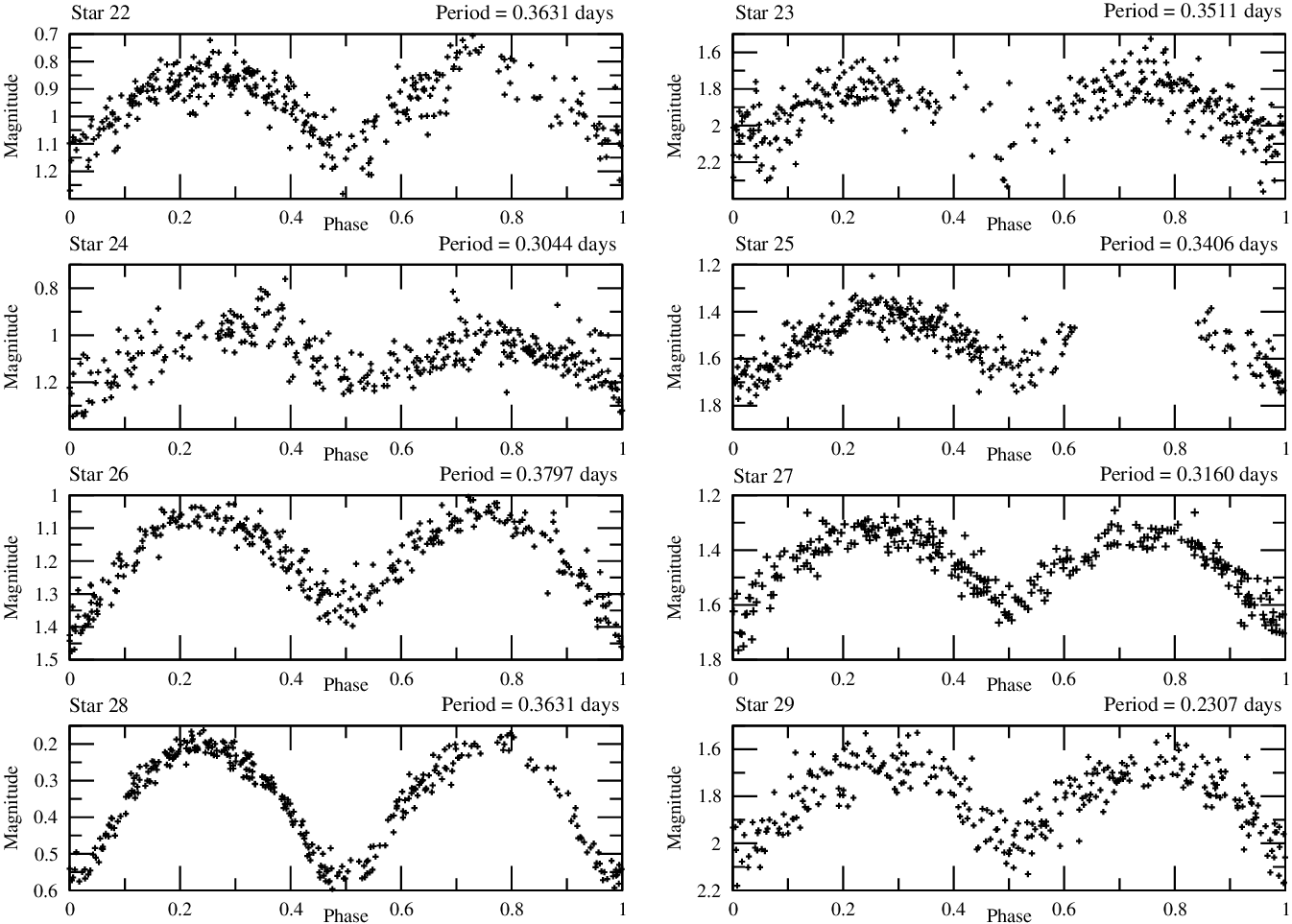}
\caption{Detected variable stars with W UMa (EW)--like light curves.\label{wuma2fig}}
\end{figure}

\begin{figure}
\plotone{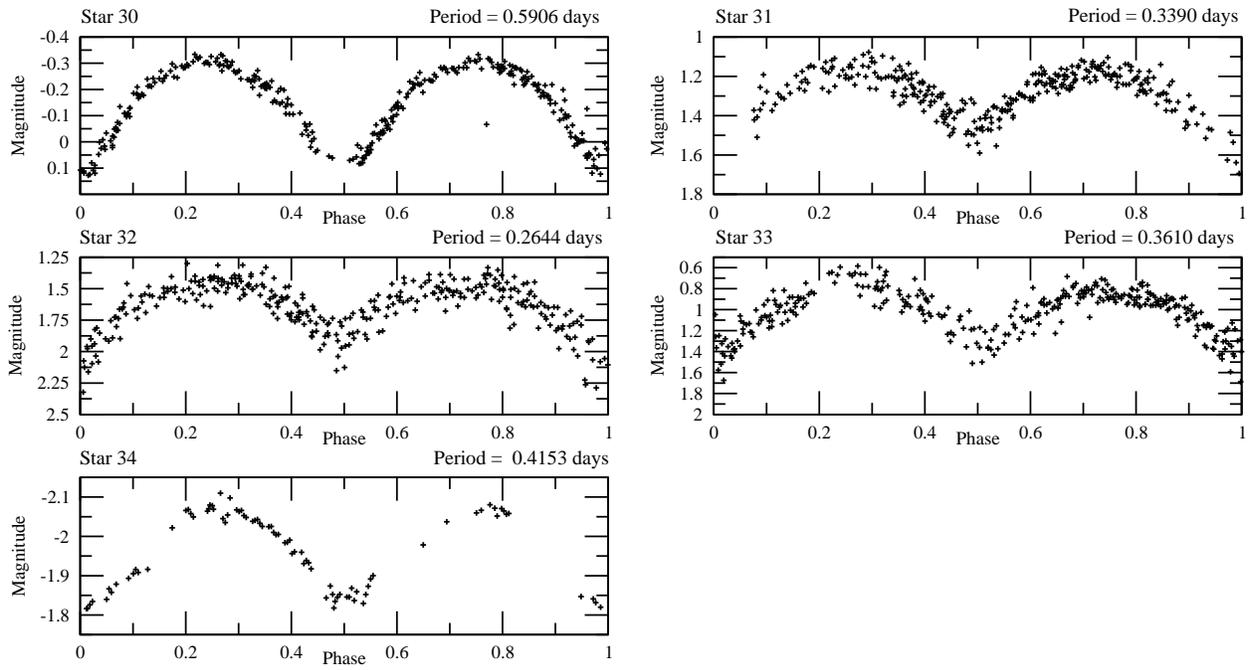}
\caption{Detected variable stars with W UMa (EW)--like light curves.\label{wuma3fig}}
\end{figure}

\begin{figure}
\plotone{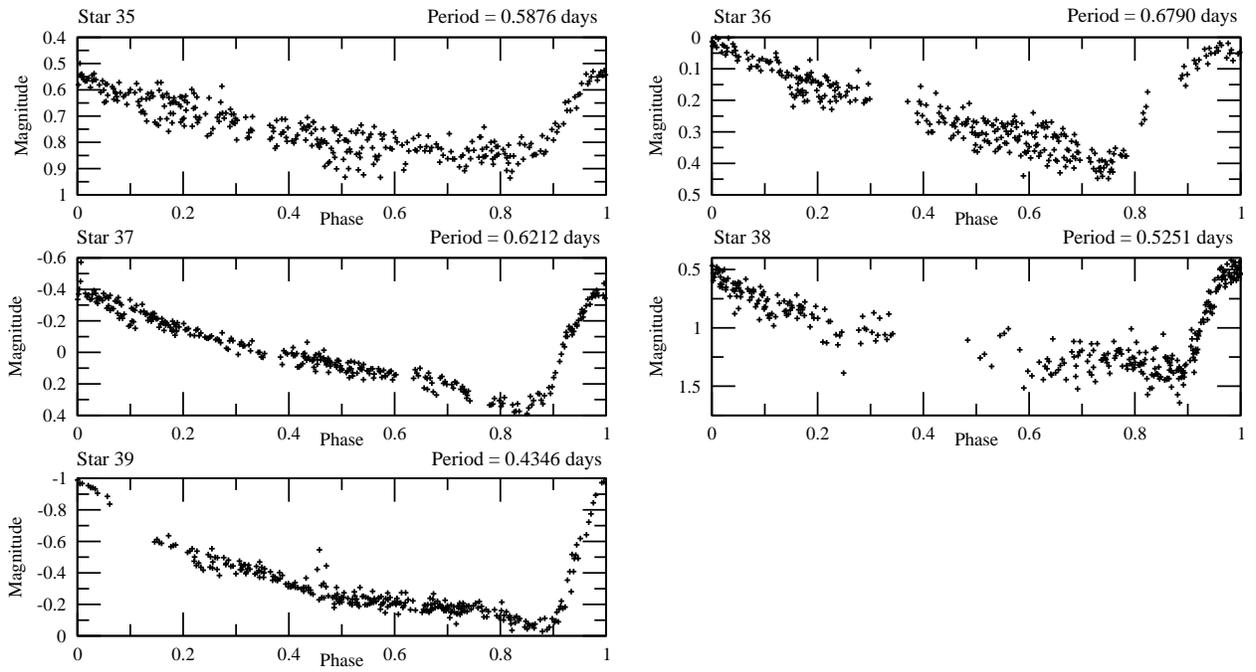}
\caption{Detected variable stars with RR Lyr light curves.\label{rrlyrfig}}
\end{figure}

Figures~\ref{algolfig}, \ref{blyrfig}, \ref{wuma1fig}, \ref{wuma2fig}, 
\ref{wuma3fig}, and \ref{rrlyrfig} are plots of the light curves for the 
stars in which we have detected variability, collected into light curve types.
It will be noted that the variability of star RAO1-01 has not been phased, and its 
identification as an Algol light curve is tentative.  The light curves of stars RAO1
-03 and -05 lack secondary minima, and, although smaller scatter favors the periods cited,
these systems could have half the periods listed.

Table~\ref{knownlist} contains a list of the known variable stars in the search 
area, detected as periodic variables with the aid of the SIMBAD and AAVSO databases.  The 
last column gives the RAO star number with which we can associate eleven of our detected 
periodic variables stars: RAO1-06, -07, -12, -15, -18, -21, -30, -34, -37, 38, and -39.  
In addition, two other variables we detected (-02 and -04) are in the vicinities of two known 
variables.  This may account for some if not all the variation apparently superimposed on 
the eclipsing light curves of these two objects.  Table~\ref{knownlist} does not contain 
stars that were over-exposed in our images except for two that lay close by the variables 
that we detected.  Of the 23 variables stars in the list that were not over-exposed, we 
detected 20 of them, and resolved the variation in 16 of these.  The variability in the 
remainder was masked by the noise sources discussed above.

\begin{figure}
\plotone{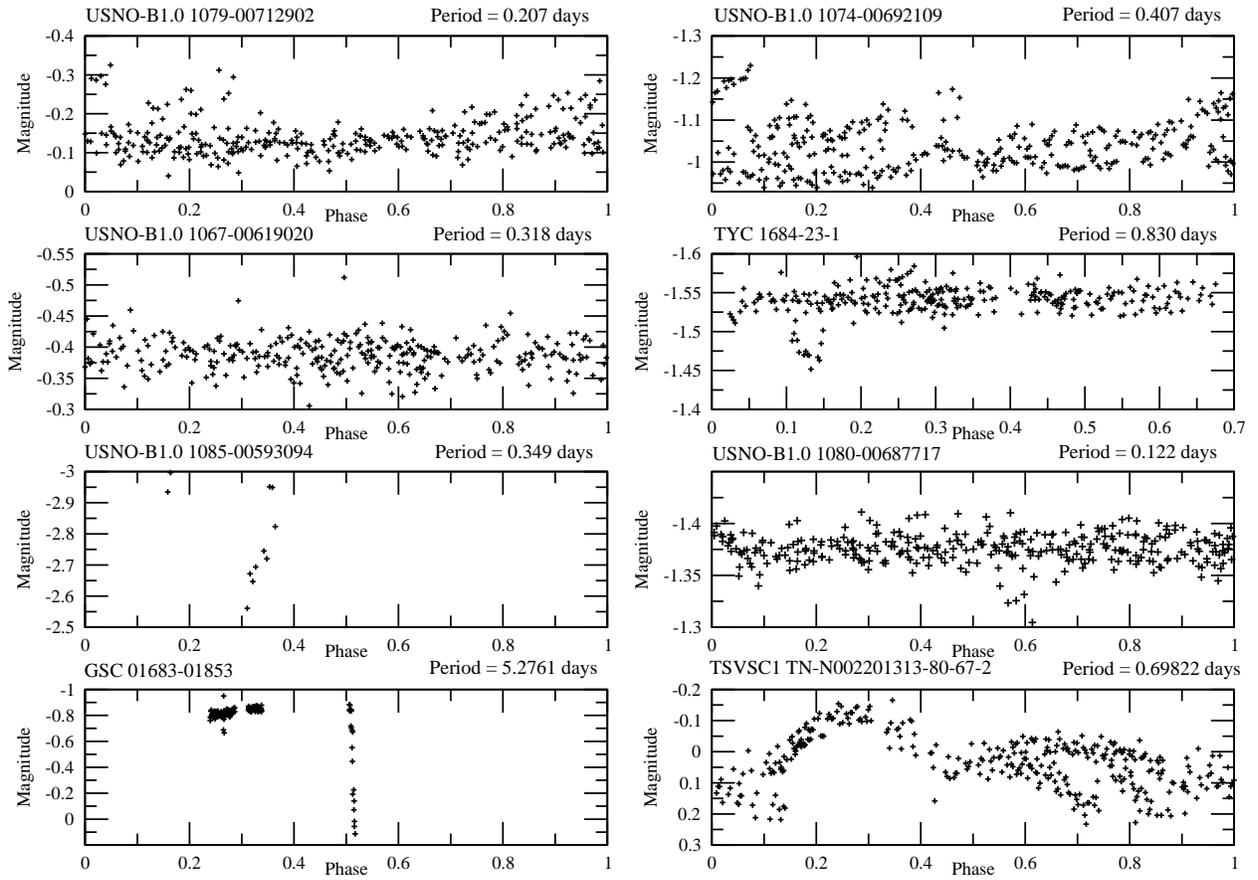}
\caption{Light curves for the previously known variable stars that were not detected as periodic
variables in our search.  The light curves are phased with the literature period.
\label{nondetectedstars01}}
\end{figure}

\begin{figure}
\plotone{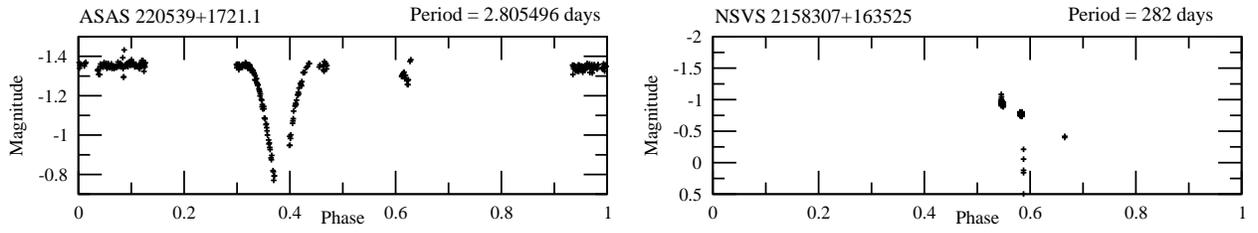}
\caption{Light curves for the previously known variable stars that were not detected as periodic 
variables in our search.\label{nondetectedstars02}}
\end{figure}

Figures~\ref{nondetectedstars01} and \ref{nondetectedstars02} show the light curves from 
our data (phased to the 
period given in the literature) for the ten previously known variable stars that we failed 
to detect as having periodic variability.  In fact, all but three of these systems we classify as 
\textit{possible} variables (marked ``DP'' in Table~\ref{knownlist}) but the systematic 
noise in the light curves precluded reliable detection of any significant degree of periodicity.  
For example, stars TYC 1684-23-1 and GSC 01683-01853 exhibit significant variation, but 
not complete cycles in the present dataset.  
Many known variable stars are within the RAO1 field but nearly all are undetected 
because the images are saturated; in some cases the star may appear too bright due to the 
blends with bright objects in close proximity.
For instance, for the ``DN'' case USNO-B1.0 1085-00593094, there are insufficient data obtained 
under good seeing conditions with which to do a period search because of the star's proximity to 
an over--exposed star.   
NSVS 11679488 (2158307+163525) is classified as ``M'', and is a long--period variable; 
this object was detected and Figure \ref{nondetectedstars01} shows some evidence of variation 
over the interval of our observations ($\sim{0.1}^P$), but hardly sufficient to see periodicity.
USNO-B1.0 1080-00687717 (12.4 DSCT, ``DN'' in our table) does not show any coherent variation at 
the published period and should be reobserved carefully.

TSVSC1 TN002201313-80-67-2 has a clearly visible phase variation for part of a cycle, but is
swamped by noise in the rest.  
Tycho 1683-00877-1 (11.40 EW, marked ``N'') we have not detected, although its position places it near RAO1-04; 
it is at the bright limit, so is likely saturated in our image frames.
ASAS 220539+1721.1 (12.40 ED, marked ``DP'') was not detected to be variable, although we expected 
it to be.  In these cases the threshold for variation was set too high to avoid false detections dur
to variability in the stellar profiles.

One of the known systems which is overexposed is the extrasolar planet 
transiting system HD~209458, a system coincidentally studied by one of us
for an MSc degree (Williams 2001)\nocite{Wil01}.

Although the statistics of small numbers must be borne in mind,
it is of interest to compare the relative frequencies of the types of variable 
stars we detected with those known from past compilations. 
For this purpose, we use primarily the nomenclature of the {\it General Catalog of Variable 
Stars}, and augmented by that in use at the {\it SVX} archive of the {\it AAVSO} website. 
From this source, the relative frequencies of short--period pulsating
variables, and the EA/ED, EB, and EW/EC types of eclipsing variables should be:
69.5, 13.7, 8.1, and 8.6\% of the total number of variables in these categories, so
in this source we find the ratio of eclipsing to pulsating systems to be 0.44.
However, we found also from the SVX site, the ratio of numbers of 
eclipsing systems (EA, EB, EW, EC, ED, and more general E categories) to
short--period pulsating stars (DCEP, RR, RRA, RRAB, RRC, RRD, DSCT, and more general 
categories), to be 1.13.  The ratio of numbers of the eclipsing binaries in the EA, EB, 
EC, ED, and EW subcategories alone) to only those in the RR Lyrae subtypes is 0.74.  
Yet we have classified 30 of the 35 detected variables to be eclipsing, so the ratio is 6:1.
These comparisons suggest that in our preliminary survey results we are undersampling 
the number of pulsating stars. 

RR Lyrae variables are more widely distributed in the galaxy, but, as giants, are more 
luminous, and so visible over a greater volume, than are the pairs of dwarfs making up 
overcontact systems, especially the most numerous of these systems, those 
with cool components.  Typically, RR Lyraes have periods between 0.4 and 0.6 day but some 
subtypes peak at shorter periods; delta Scuti stars may have much shorter periods, whereas 
W UMa systems typically have periods between 0.2 and 0.4 day; systems with $\beta$ Lyrae
light curves have a greater range of periods, and Algol-like light curve systems, an
even greater range. 

We now consider it unlikely that we have misidentified many pulsating light curves,
especially those of the DSCT (delta Scuti) stars, as W UMa light curves, but some
misidentifications are certainly possible, as is multiple variability.  In the
case of misidentification, the pulsating star's
period would be half that given in Table~\ref{knownlist}.  We note that several
systems identified here as of EW type have nearly sinusoidal light curves and a number
of these have shallow light curves.  In an EW system, a low amplitude light curve
implies low inclination, whereas in delta Scuti systems, low amplitudes are
common.  However, DSCT stars constitute 4.5\% of all pulsating stars in the SVX survey
and only 2.1\% of the combination of short-period variables in the
the pulsating and eclipsing variable star groups.  In principle, other of our EW systems 
could be members of the subclass (RRC) of RR Lyrae stars; 
RRC stars have sinusoidal light curves with lower amplitudes (typically 0.5 magn) and
constitute 10.2\% of RR Lyrae type variables.  
Color information as well as spectroscopy can be used to discriminate between EW 
and RRC types, as pointed out some time ago by Wesselink et al. (1976)\nocite{Wes76}.  

The ellipsoidals are yet another class of variable with low-amplitude sinusoidal light curves.
These are even rarer objects, however, making up less than 0.3\% of short--period variables. 
Finally, some low-amplitude objects may be spotted, and relatively fast rotators.  Where this is 
the case, the amplitudes and shapes should vary over cycles; low--amplitude 
candidates in our sample include stars RAO1-14, -16, and -20 (three of the 19 we intially 
classified as EWs) --- all potential members of this group.

Three possiblities for the paucity of RR Lyrae stars in our data set are that:
\begin{enumerate}
\item{There is a bias due to differences in spatial distribution.
Objects of the old disk population, to which most of the dwarf binary systems likely belong,
are more concentrated near the galactic equator than are population II RR Lyraes, which
are more spherically distributed;}
\item{There is a luminosity bias in the survey because of our effective magnitude range.  
The more luminous RR Lyrae stars that are close to us are overexposed, whereas the more 
distant ones are preferentially dimmed and reddened as a consequence of being more distant 
and thus of having greater interstellar extinction than do the nearby overcontact binary star components;}
\item{There is a bias toward shorter period systems in our survey, favoring W UMa systems, for
example, over the statistically slightly longer period RR Lyr systems.}
\end{enumerate}

The mean brightness of an RR Lyr star is $M_V \approx{0.7}$. 
The distant modulus range that we sample within the bright and faint magnitude limits of 
our survey is $10.6 \le (V-M_V) \le 16.3$ for the RR stars, and $7.2 \le (V - M_V) \le 12.9$ 
for a pair of solar--type stars, and for a mixed population of binaries over a range of 5 
magnitudes of $M_V$, centered on $M_V$ = 6,  $\sim{3} \le (V - M_V \le \sim{14}$.  Interstellar 
extinction makes objects appear fainter and thus more distant, but the true volume is not plumbed 
more deeply, so that the number of stars is not thereby increased.

Regarding hypothesis 1, the galactic coordinates of RAO field 1 ($l_{II}$ = 76.8$\arcdeg$, 
$b_{II}$ = 28.5$\arcdeg$) place it $\sim{30}\arcdeg$ off the galactic plane, so that the 
frequency of old disk and younger objects relative to that of the RR Lyrae types should be
decreased compared to the ratio in the galactic plane; howeve it will be large.  
The mainly dwarf type stars that constitute most of the binaries we detect can not be seen 
at distances as great as the RR stars. The slant direction thickness through the galaxy's thick 
disk is double the path length through the normal to the galactic plane, and this certainly 
increases the numbers of detected dwarfs relative to RR stars.  Therefore the hypothesis
must be correct at some level, but it is hardly clear that it is quantitatively sufficient.

Hypothesis 2 also has merit because even with minimal extinction, the brightest RR Lyrae star
that we can detect must be more than a kiloparsec away, if interstellar extinction is less than 0.5 magn,
although there is a dependence on metallicity (Hall 2000, p.400)\nocite{Hall00}. 
If we take into account that the RR Lyrae stars are themselves not uniform in spatial,
metallicity, or period distributions, the argument become stronger.
Classically there is a dichotomy in the RR Lyrae group (Gilmore \& Zeilik 2000, p. 479)
\nocite{GilZei00} between RR stars with periods greater than $0.4^d$ for which the galactic 
scale height, $h$, is 2kpc, and those those with periods less than $0.4^d$, for which $h=300$ pc.
These numbers can be compared with the scale height for old population I stars, 100 pc.
The shorter period RR stars are said to be more numerous in the solar neighborhood,
and these would be brighter as a consequence, so bright that they will exceed our bright limit.
Because these stars are saturated on images, so they will not be measured and will not get into the 
database.  The dust is confined primarily to the galactic plane, and the extinction in the plane is 
typically $\sim{1}~magn/Kpc$.  Much of the extinction must be local, but the net result will
be to decrease the numbers of the more distant stars; indeed, because the sample of
stars is actually drawn from a segment of the celestial sphere with a smaller distance, the
volume sampled will be smaller.  Therefore this 
hypothesis, too, appears necessary but may not be sufficient to explain the full disparity.

Hypothesis 3 certainly applies, because the shorter-period RR's are distributed closer to
the plane, and these are too bright to be seen.  All five of our detected RR stars have
periods longer than $0.4^d$ (see Fig. 10).
Thus it appears that all the hypotheses have merit because they involve closely correlated
properties of the RR Lyrae stars.  Again, the last hypothsis is necessary if not sufficient
to explain the relative paucity of RR stars in our discovery list.
The qualitative explanation may be correct, but a larger body of data will permit tests of the
hypotheses.  To that end, we are acquiring images with a range of exposure times,
thus expanding the dynamic range of the brightness, and distance, of detected variables.

Looking now at the major subgroups in more detail, we see that among the EA, EB, and EW/EC cases 
in the SVX database, the percentages of these types are
33.4, 5.6, and 60.9\%, respectively.  This can be compared with 20, 17, and 63\%, 
respectively, of our sample of the 30 independent variables detected according to our 
criteria.  Therefore the number of over--contact systems we detect is as expected, whereas 
the Algols appear under-represented, and the $\beta$ Lyraes, overabundant, notwithstanding 
the statistical uncertainty in these small numbers.  A relative paucity of Algol (EA) systems 
relative to W UMa (EW) systems, though, is not difficult to explain.  The relatively short 
interval of observations and short durations of minima of the EA systems combine to limit the 
number of such systems that can be detected in our data.  The distribution of periods 
is far greater for the Algols than for the W UMa types, so over short intervals the 
long--period Algols will be missed.  
Thus the discrepancy between EAs and EWs is readily explained, at least qualitatively.

Finally, this discussion has been about \textit{light curve} as opposed to \textit{binary} morphology.
It is another matter entirely to identify systems not according to the shape of their light 
curves, but according to their state of binary evolution.  The characteristic
rounded maxima that signals a star's departure from spherical shape is usually absent from the
visual light curves of systems with Algol--like light curves: the hotter
component is more luminous than its low--surface brightness companion, and dominates optical
light curves.  In infrared light curves, where a larger contribution comes from the larger,
cooler component, the curvature is observed; in the IR, Algol itself is seen to have rounded maxima. 

To identify the morphological types, and
to aid in any further light curve studies, we have performed preliminary analyses of the 
detected eclipsing systems for which both minima are observed.  The converged solutions
may be useful as starting values for analyses of larger and more precise data sets,
and to identify particularly interesting targets.

\subsection{Preliminary Light Curve Analyses}

Analyses were attempted on all the eclipsing binaries in Table~\ref{starlist} 
by EFM with the WD package described initially by Kallrath et al. (1998)\nocite{Kaletal98}, and
updated to include improvements up to 2007 (see Milone \& Kallrath 2008\nocite{MilKal08}).  
The package contains a simplex program, a Wilson--Devinney program (version $\sim{1998}$), 
a damped least--squares (\texttt{DLS}) version of the latter program, and a simulated annealing 
(\texttt{SA}) program, which, however, was used only sparingly in the present analyses.  
The SA program, described and illustrated in Milone \& Kallrath (2008)\nocite{MilKal08}, is as 
robust a light curve solver as one may find, but it is not as efficient as the Simplex program at 
finding global solutions, and in the present application where a large number of light curves had 
to be analyzed, a starting run with the simplex routine usually sufficed.

In the present study, the simplex algorithm was run, preceding a self--iterating version of
the Wilson--Devinney program; on completion, a final DC and LC run then gave the least-squares
parameter values and their mean standard errors (mses).  Either the simplex or the DLS iterations 
ended when either full convergence was obtained, i.e., when the correction for each parameter 
became less that its mse, or when one of the stopping criteria was reached. Stopping criteria 
include preset values of the number of iterations, the number of changes of the Marquardt damping 
factor, or an error limit, typically selected as the uncertainty in a single observational datum. 
The latter is compared to the mse of the fitting.  The stopping criteria 
are stipulated in an information file, along with the programs to be used in
sequence, the beginning CGS {\it log~g} values for each component (automatically updated as 
the model changes during the calculation), the name of the flux file that contains the ratio of 
the atmospheric flux (from the models of Kurucz (1993)\nocite{Kurucz93} to that of a blackbody 
(in the present case, only that for the $R_J$ passband), and a number of other control parameters.
The Differential Corrections (\texttt{DC}) input file and a constraints file which lists the 
ranges over which parameters may be varied to keep the exploration within realistic physical limits, 
are also required to run the package.

We typically started each system's run in mode 2, in the parlance of the Wilson--Devinney program,
which is appropriate for fully detached systems.  
These modes (and the models for which they are appropriate) are: 2 (detached system), 3 
(overcontact), 4,5 (semi-detached with primary, secondary filling its Roche Lobe, respectively), 
and 6, for contact systems, where both stars fill their inner lobes and are thus in marginal 
contact.  It is standard among light curve modelers to determine the proximity of the computed 
stellar radius to its inner Lagrangian surface, or Roche lobe, with the \textit{fill-out 
parameter}, defined by
\begin{equation}
\label{filloutequ}
f = \frac{\Omega^{I} - \Omega}{\Omega^{I} - \Omega^{O}},
\end{equation}
where $\Omega$ is the modified Kopal potential and $\Omega^{I}$ and $\Omega^{O}$ are the inner 
and outer Lagrangian surface potentials, respectively. Where f $\approx{0}$, the component is 
near its inner Lagrangian surface, and where f $>$ 0, it exceeds this limit.  This quantity is 
computed in the simplex program output within the WD package.

If convergence was not achieved in mode 2 initially, we made further runs with the best 
values of the preceding run as starting parameters, with a recomputed semi--major axis from the 
new period value.  At this point, the data set was reexamined and residuals that exceeded the 
Chauvenet criterion were given zero weight (although they still appear in the final fitting 
plots).  If, after several such attempts, convergence was not achieved in mode 2, or if the 
fill--out parameter indicated that the components were close to or exceeded their Roche lobes, 
a run was made in the over--contact mode.  Finally, when no convergence was found after several 
such attempts, or the fill--out parameters suggested them to be viable models, a run was made
in modes 4, 5, or 6, as appeared appropriate.

In all the cases of the light curves identified as ``WUMa''  (``EW'') type, we have 
explored fittings with both detached (mode 2) as well as over-contact models (mode 3), and where
f $\approx{0}$, for either or both components, we modeled the systems also in
modes 4, 5, or 6.

In light curves where the maxima are curved, we attempted to model the photometric
mass ratio among the full suite of parameters.  In other cases, we modeled only the inclination, 
$i$, the temperature of the secondary
component, $T_2$, the modified Kopal potential $\Omega_1$, the period, $P$, the epoch, $t_0$, and
the passband luminosity, $L_1$.  For detached systems, we adjusted also the potential of the
secondary component, $\Omega_2$.  We employ here the photometric definition of ``primary'' and 
``secondary'' stars as those at superior conjunction at the primary (deeper) and secondary 
eclipse, respectively.

Because we have no color or spectral information about the primary component, in almost all 
cases, we arbitrarily set $T_1$ = 6000 K.  The lone exception was RAO1-29, a faint, very 
short--period system, for which we set $T_1$ = 5000 K. 

In all cases, the starting period and epoch are those derived by MDW and which appear in Table 3,
in the ``Final Period'' column.

The results of the analyses for 26 converged cases (iincluding two for RAO1-06, and one each for 
the duplicate pair RAO1-10 and RAO1-11) are listed in Table ~\ref{solutionslist}.
The fittings and residuals of the fits are shown in Figures \ref{solutions1} -- \ref{solutions9}.
The number of observations in the light curve appears in column 2 and the mode of the adopted 
model appears in column 3.  

Although $T_2$ is listed as a parameter in the output lists of the DC program, in effect
the program calculates only the difference in temperature between the two components, as 
determined primarily by the difference in depth of the two minima.  Thus we list only the 
temperature difference, $\Delta$T = $T_2$ - $T_1$, and its uncertainty, in column 5.  

The last column contains the standard deviation of a single observation of average weight.
The other columns contain the parameters and their mse uncertainties.  Where no uncertainty is
given, the quantity was computed internally from other quantities or was assumed.

From the 25 converged solutions, we can categorize the systems: 10 detached, 3 semi--detached,
10 overcontact, and 2 contact systems. However, the detached systems RAO1-10 and -11 are very 
likely contain information about the same variable star, but contain the light of another object 
in either greater or less proportion, so that we actually have 24 independent systems (9 detached 
systems) for which solutions are found.  Over--contact systems RAO1-22 and -28, for which the 
detection process found the same period, however, have produced different periods and other 
properties in the light curve analyses, and may well be independent variables, afterall.
Next we will describe next some of the potentially most interesting systems to emerge from the analyses.

\subsection{Solutions}\label{solutions}
\begin{figure}
\plotone{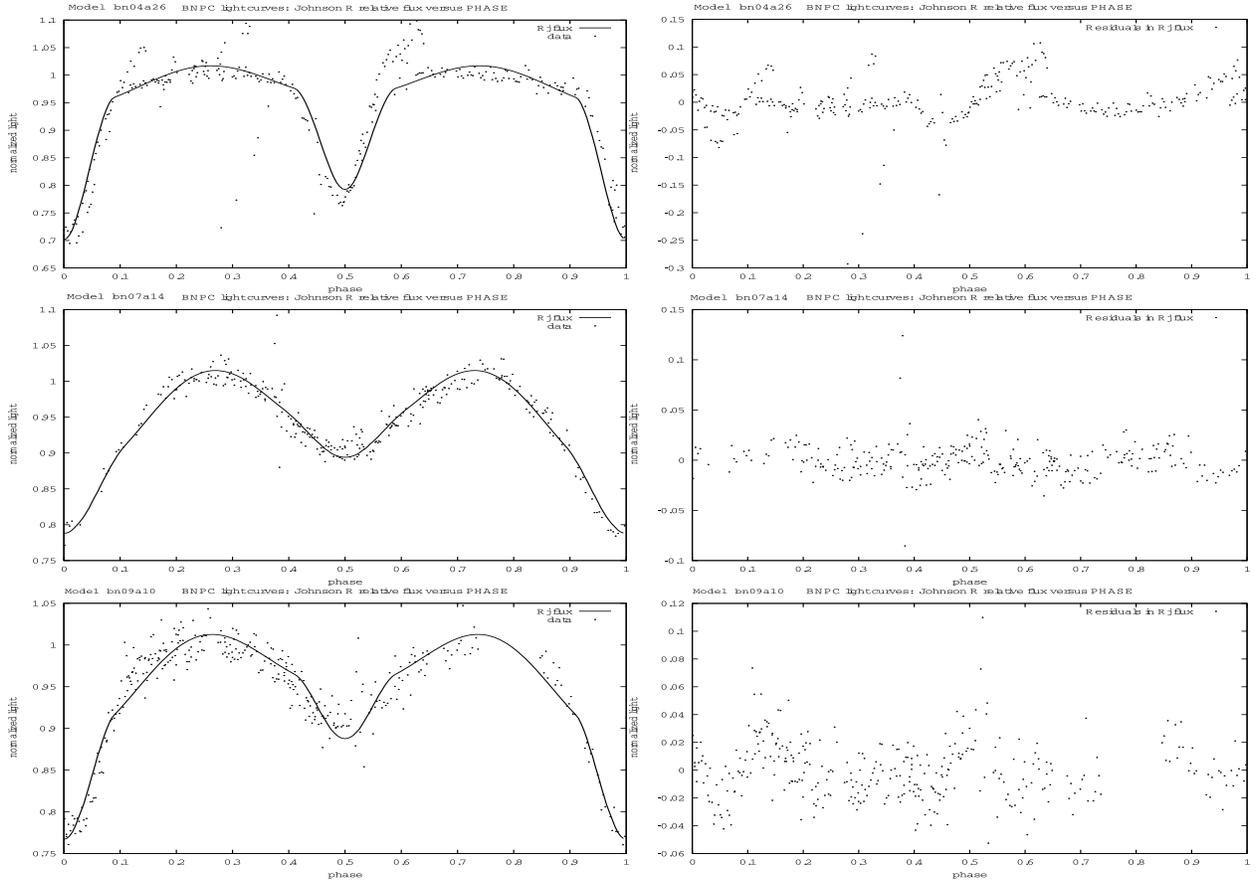}
\caption{Light curves and fittings (left) and residuals (right) for the eclipsing binaries 
for which we have obtained converged solutions: The fittings for RAO1-04, -07, and -09. 
The light curves are phased with the elements found in the light curve fittings.\label{solutions1}}
\end{figure}

\begin{figure}
\plotone{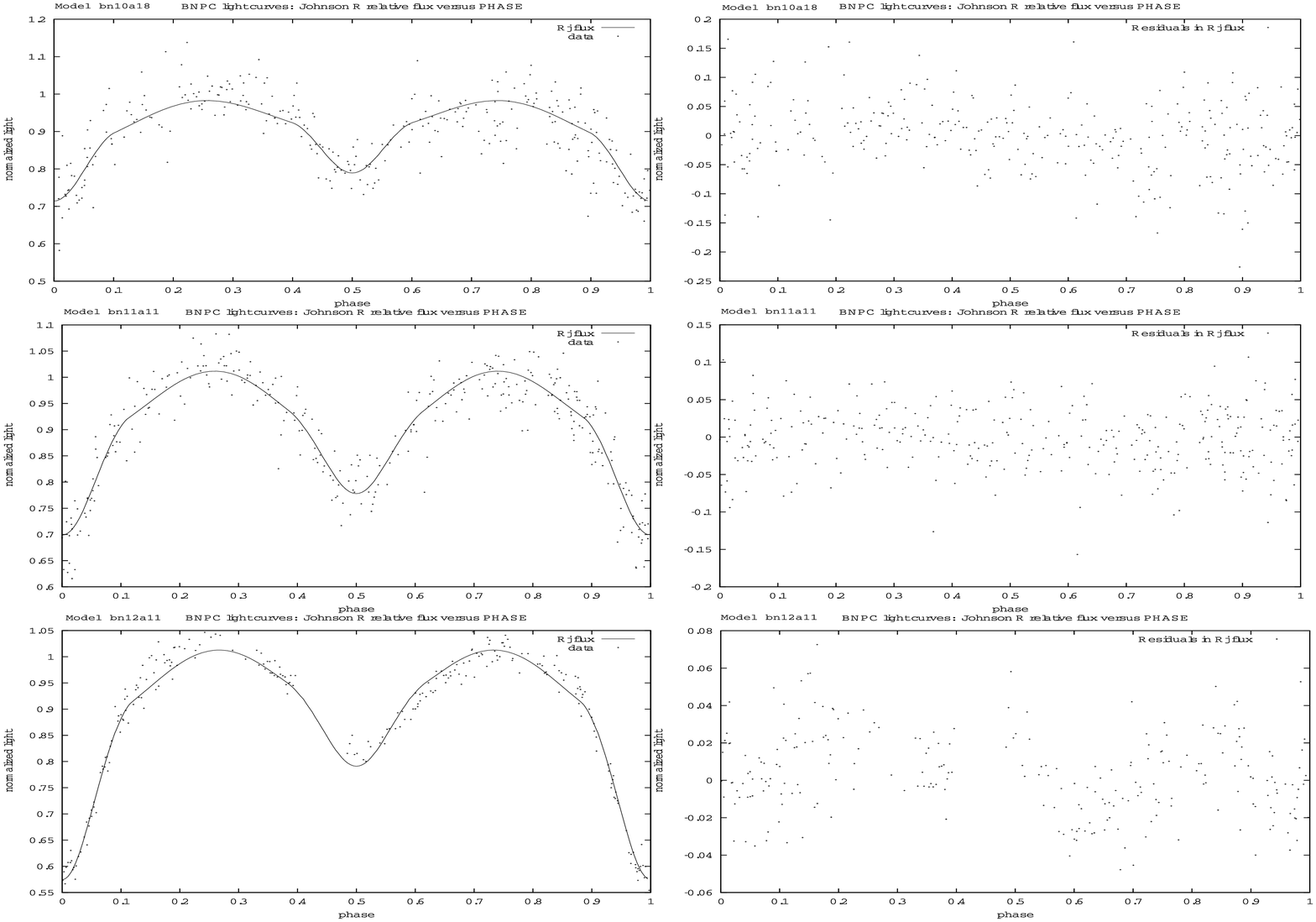}
\caption{Light curves and fittings (left) and residuals (right) for the eclipsing binaries 
for which we have obtained converged solutions: RAO1-10, -11, and 12. 
The light curves are phased with the elements found in the light curve fittings.\label{solutions2}}
\end{figure}

\begin{figure}
\plotone{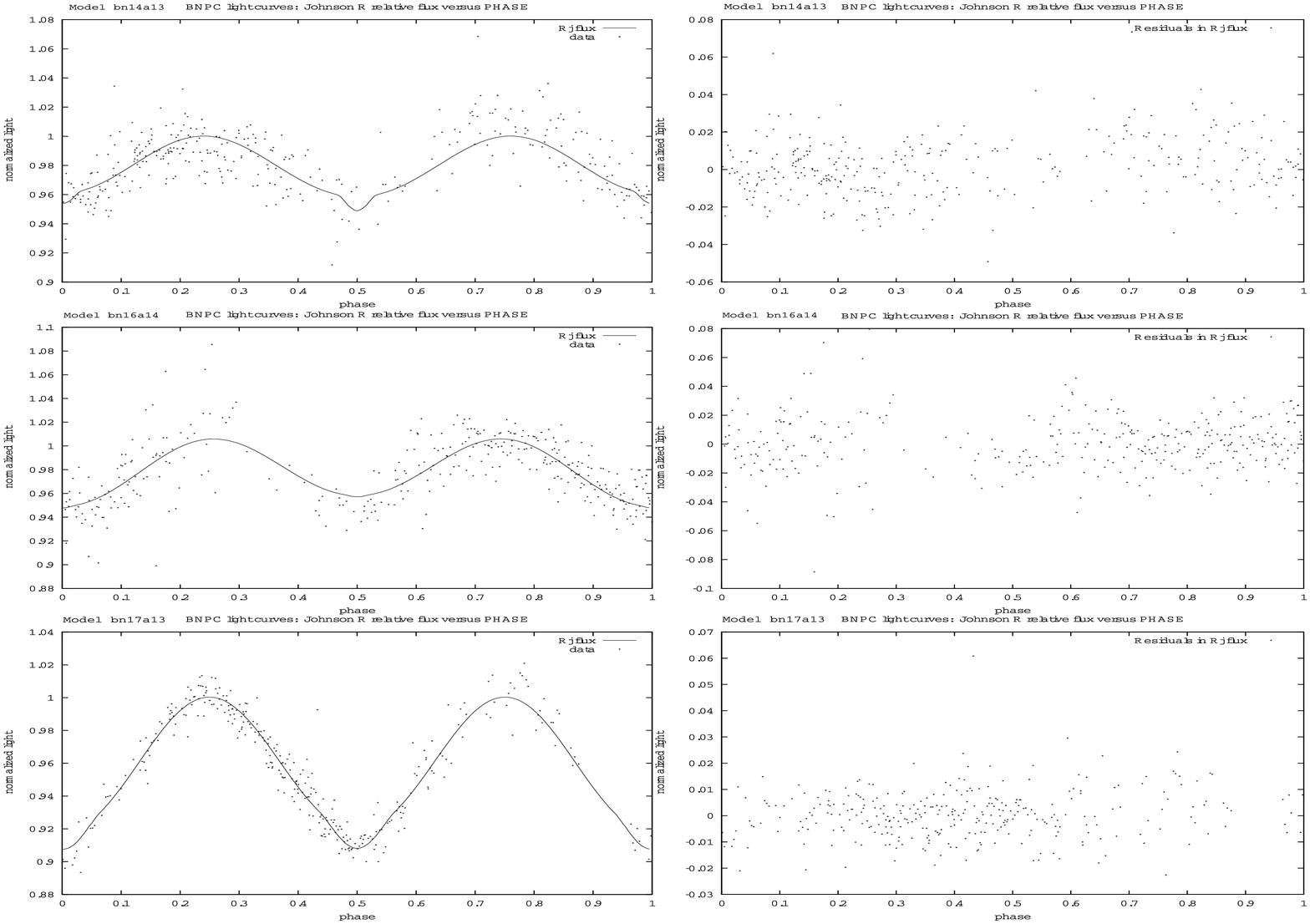}
\caption{Light curves and fittings (left) and residuals (right) for the eclipsing binaries 
for which we have obtained converged solutions: RAO1-14, -16, and -17.
The light curves are phased with the elements found in the light curve fittings.\label{solutions3}}
\end{figure}

\begin{figure}
\plotone{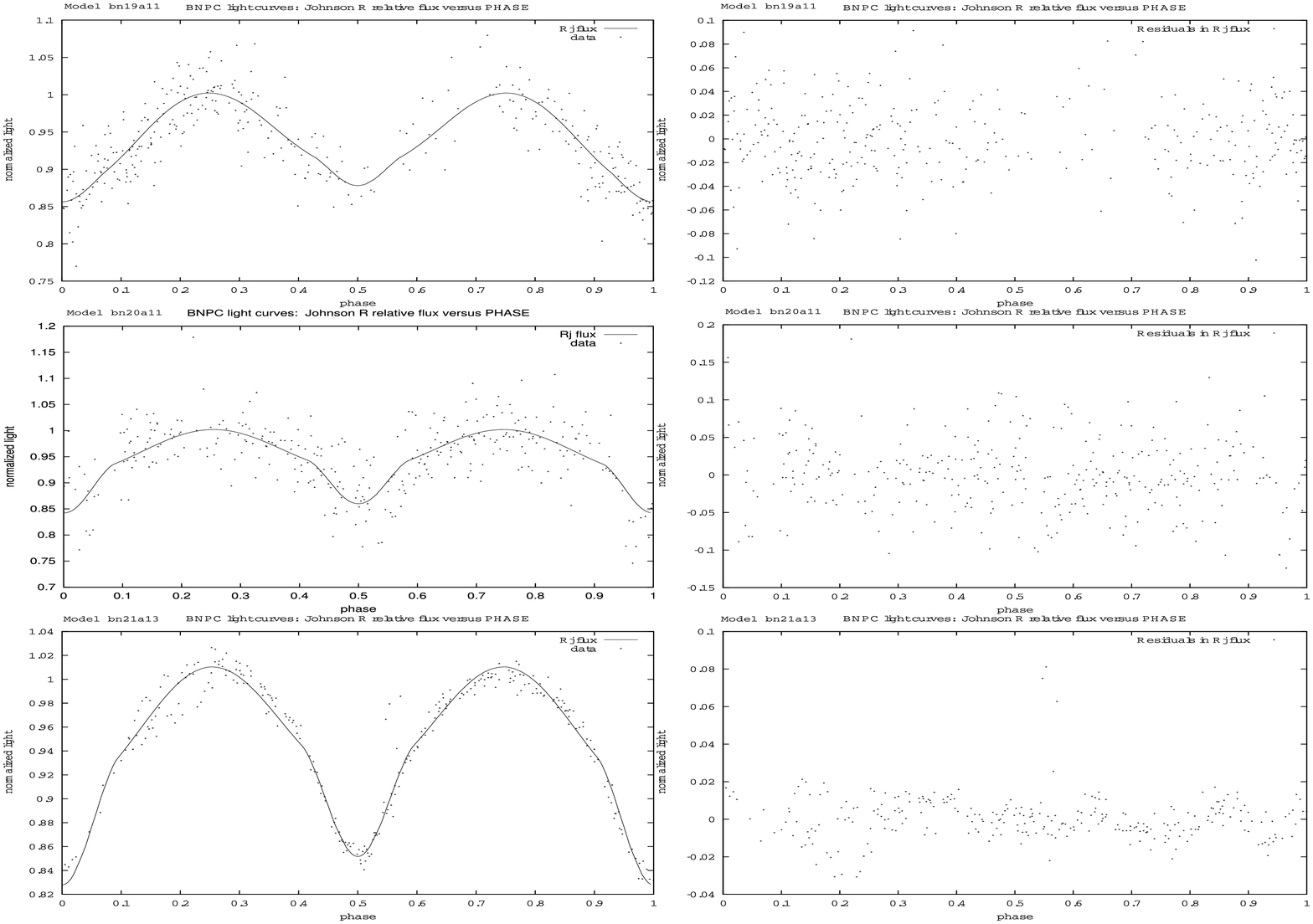}
\caption{Light curves and fittings (left) and residuals (right) for the eclipsing binaries 
for which we have obtained converged solutions: RAO1-19, -20, and -21.
The light curves are phased with the elements found in the light curve fittings.\label{solutions4}}
\end{figure}

The light curve of RAO1-04 is distorted with what appear to be variations in system light, so the
data with the worst residuals were given zero weight.  The slight curvature at maximum light makes
the direct determination of mass ratio as one of the simultaneously adjusted parameters 
problematic.  
In such a case, however,  a grid search, where only one parameter can be adjusted, can be 
performed over a range of values to find a minimum sum of squares of residuals (SSR).  
For RAO1-04, $q$ was adjusted in 20 trials across the range 0.75 $\leq$~q~$\leq$ 1.30 to 
determine the mass ratio for RAO1-04.  Optimization was achieved for q = 1.1 in this detached 
system, which shows only a slight curvature in the maxima.  The fitting and residuals plots
can be seen in Fig.~\ref{solutions1}.

One of the more interesting systems is RAO1-06, a detached system, which in the adopted model has 
residuals that appear to follow a sinusoid, indicating a possible low-amplitude variable in the 
system.  
The photometric mass ratio could not be determined in this system
from the single $R_J$ passband and single epoch light curve alone, and so was assumed to be 
1 and not adjusted. The earlier modeling trials for RAO1-06 assumed a more complicated reflection 
solution than is usually assumed in an attempt to reproduce the apparent positive slope in the 
light curve between phases 0 and 0.5.  Usually a large ``reflection effect'' occurs in systems 
with a large difference in sizes and temperatures of the components.  However, in this case, 
results of early trials indicated these differences to be relatively small, and enhanced
reflection treatment did not, in fact, diminish the residuals.  
The parameters for the best--fitting model to this point, bn06a13, is listed in 
Table~\ref{solutionslist} on the RAO1-06 line.
It was possible to reduce the effects of the residuals by subtracting the following function:
\begin{equation}
\label{06residequ}
f(res) = -0.036\times{sin~ (1.45\phi)}~ + ~0.008 
\end{equation}
where $\phi$ is the phase in radians.  
This quantity was subtracted from the light curve to create the results listed for the object RAO1-06s;
the fitting and residuals as well as the sinusoid superimposed on the residuals of model bn06a13 are
shown in Figure~\ref{solutions9}.  The eclipsing light curve is incomplete, so not too much should be
made of this sinusoid representation beyond that further observations of this system are warranted.

The systems RAO1-07 and -12 were modeled as semi--detached cases after detached and overcontact models
failed to produce convergence.  For RAO1-07, the fill--out parameters were $f_1$ = -0.18 and $f_2$ = +0.06;
the former value indicates that the primary component is within its Roche lobe, and the latter value indicates 
that the secondary component just slightly exceeds its Roche lobe. Although no solution was obtained in
mode 3, the model was not far, in fact, from convergence.  In the case of RAO1-12, the light curve
is missing part of the secondary maximum (the maximum following the secondary minimum), but the other
parts of the light curve could be well--fitted with the semi--detached model bn12a11.  A converged
solution was found also in mode 2, but the fitting was not as good as that for mode 5.  The fill--out
parameters for the mode 5 model were $f_1$ = -1.34 and $f_2$ = +0.03.  The fittings for the systems'
converged solution models, bn07a14 and bn12a11, can be seen in Figs.~\ref{solutions1} for RAO1-07,
and \ref{solutions2} for RAO1-12, respectively.

\begin{figure}
\plotone{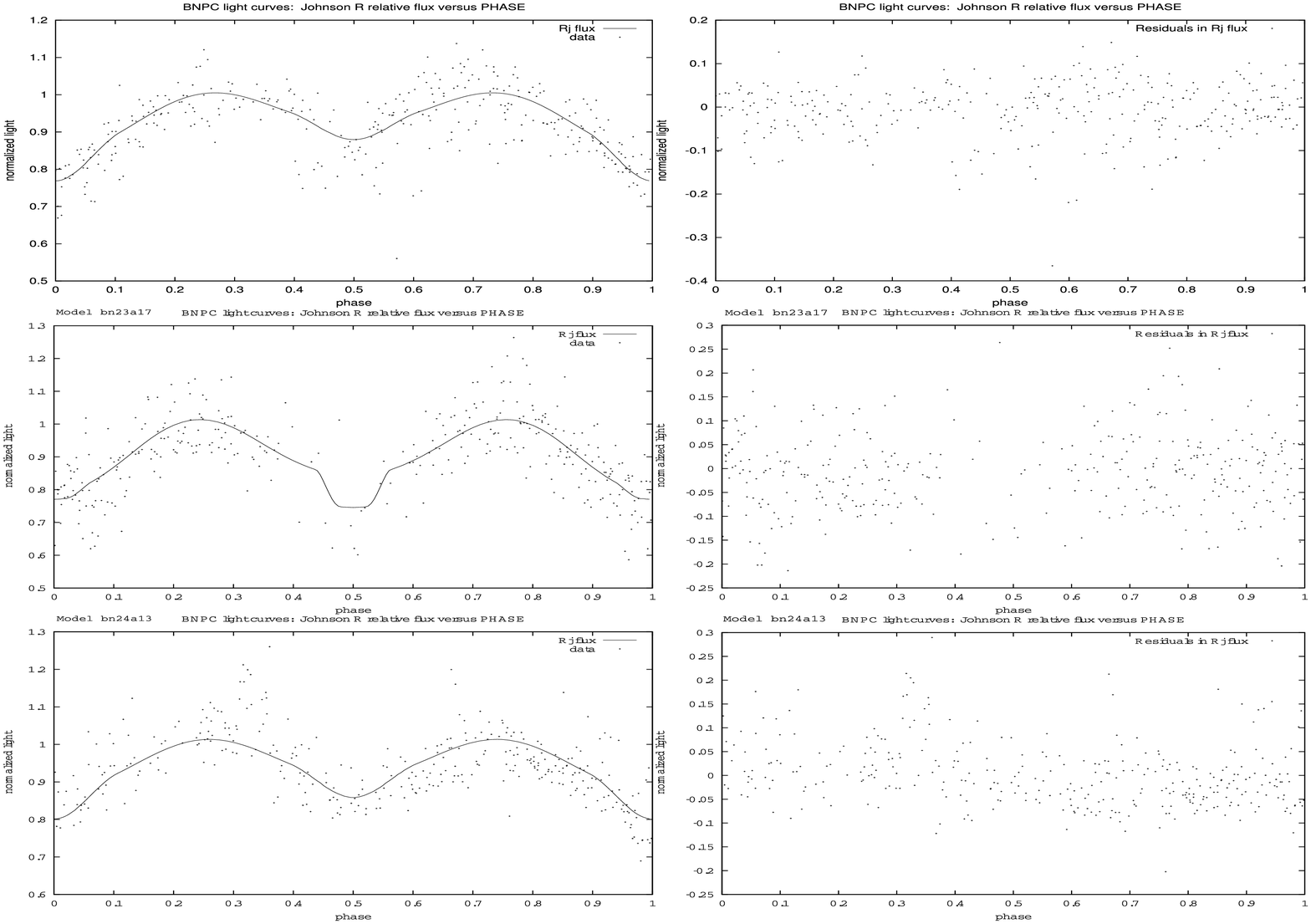}
\caption{Light curves and fittings (left) and residuals (right) for the eclipsing binaries 
for which we have obtained converged solutions: RAO1-22, -23, and -24.
The light curves are phased with the elements found in the light curve fittings.\label{solutions5}}
\end{figure}

\begin{figure}
\plotone{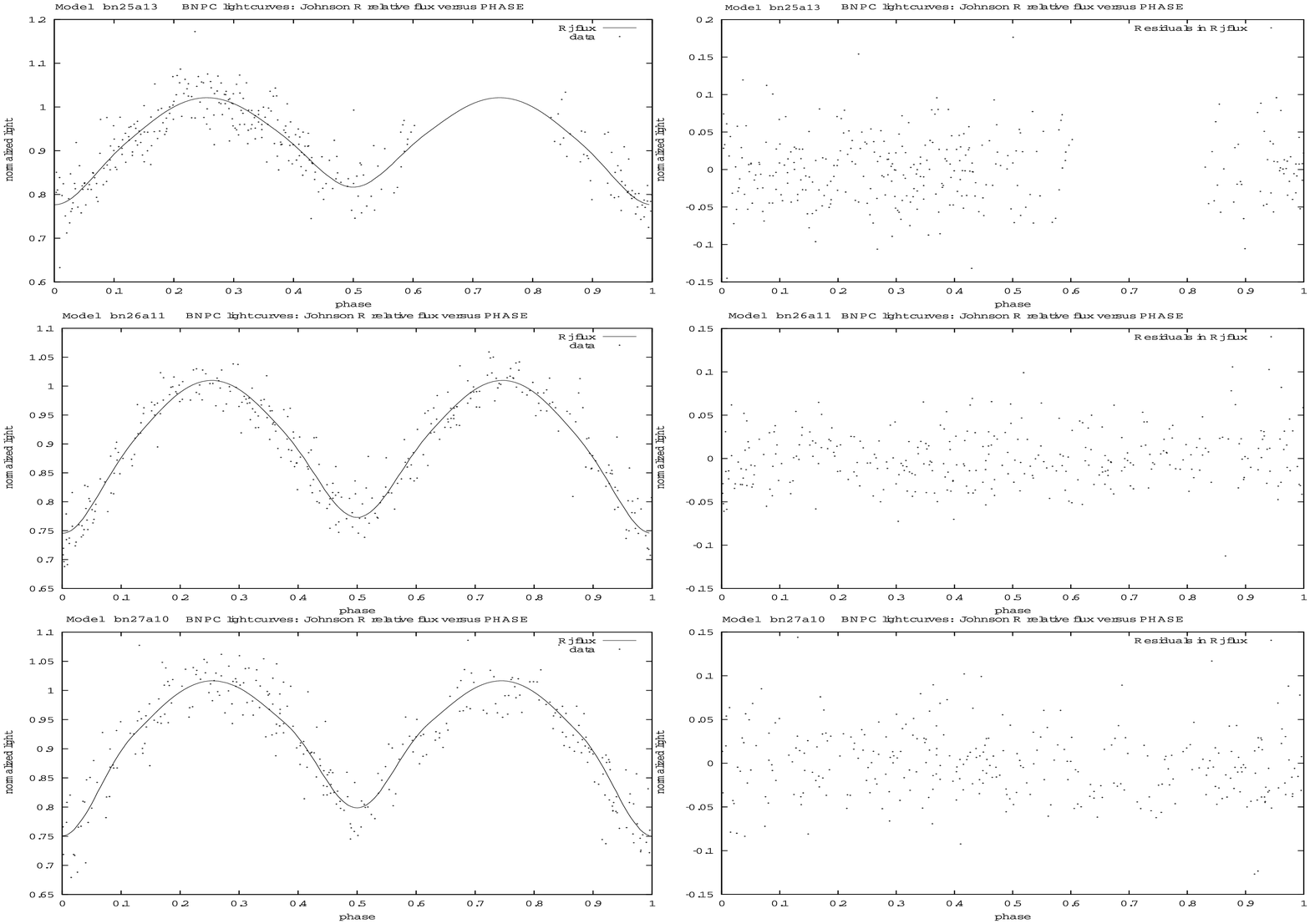}
\caption{Light curves and fittings (left) and residuals (right) for the eclipsing binaries 
for which we have obtained converged solutions: RAO1-25, -26, and -27.
The light curves are phased with the elements found in the light curve fittings.\label{solutions6}}
\end{figure}

\begin{figure}
\plotone{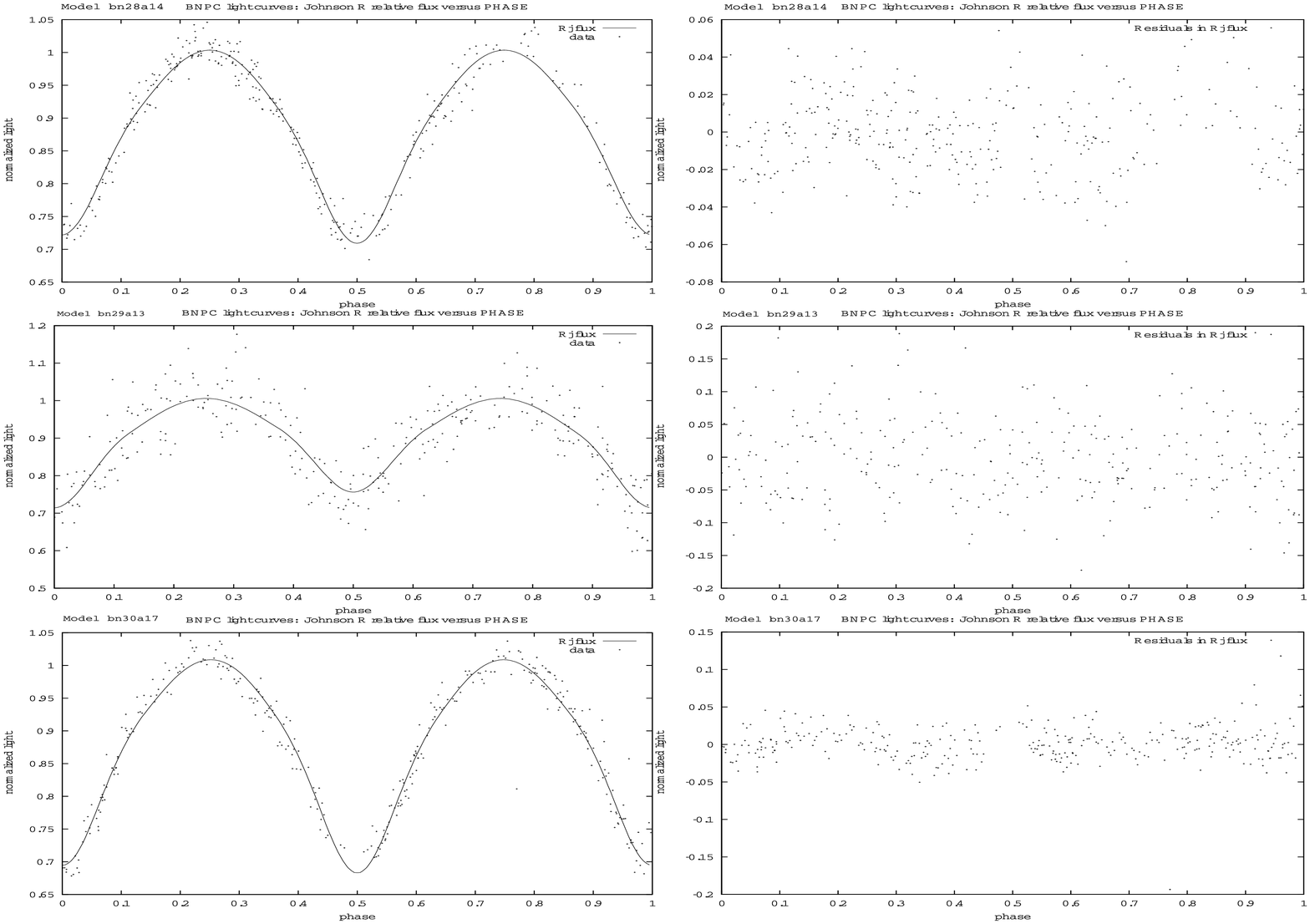}
\caption{Light curves and fittings (left) and residuals (right) for the eclipsing binaries 
for which we have obtained converged solutions: RAO1-28, -29, and -30.
The light curves are phased with the elements found in the light curve fittings.\label{solutions7}}
\end{figure}

\begin{figure}
\plotone{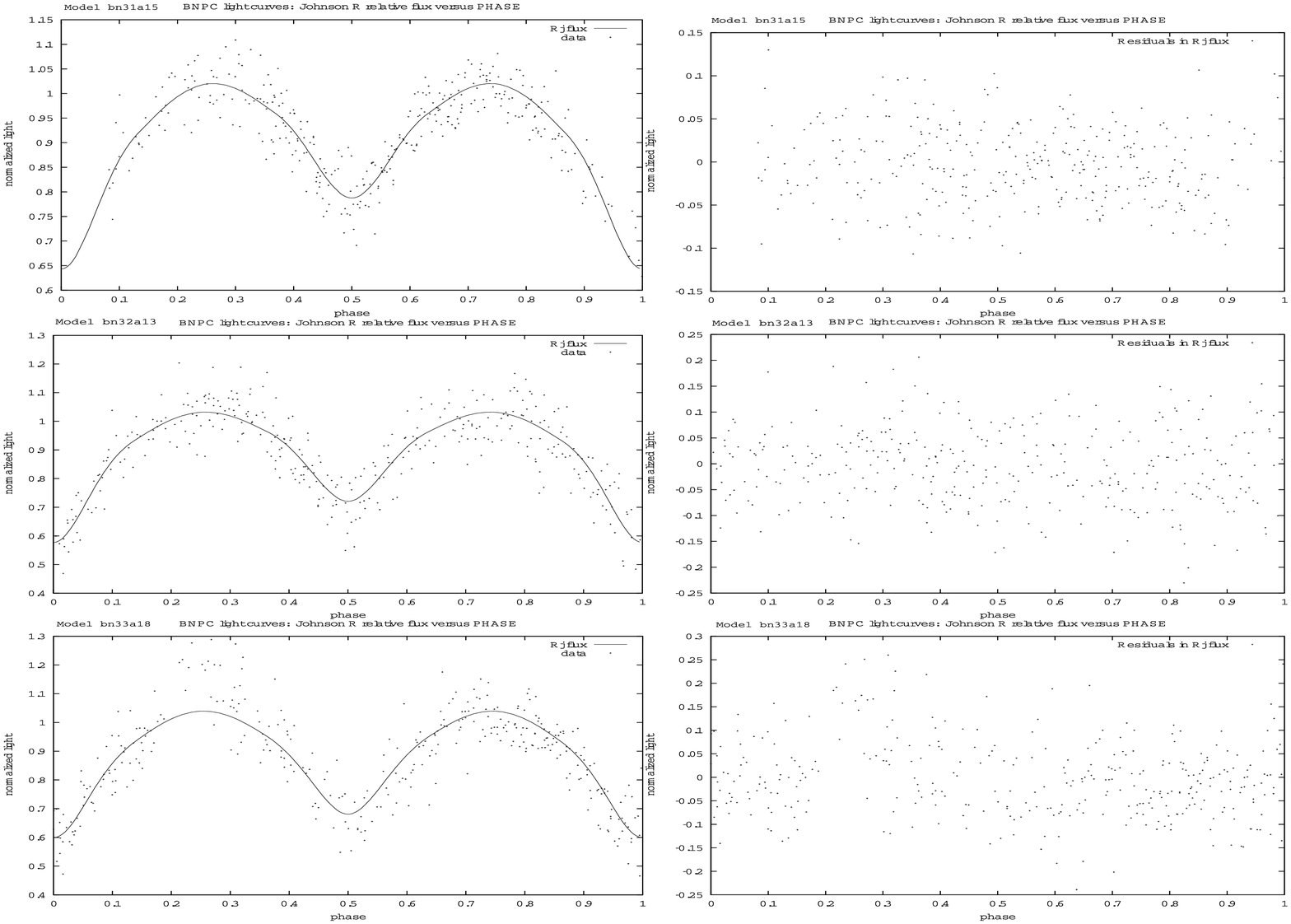}
\caption{Light curves and fittings (left) and residuals (right) for the eclipsing binaries 
for which we have obtained converged solutions: RAO1-31, -32, and -33.
The light curves are phased with the elements found in the light curve fittings.\label{solutions8}}
\end{figure}

\begin{figure}
\plotone{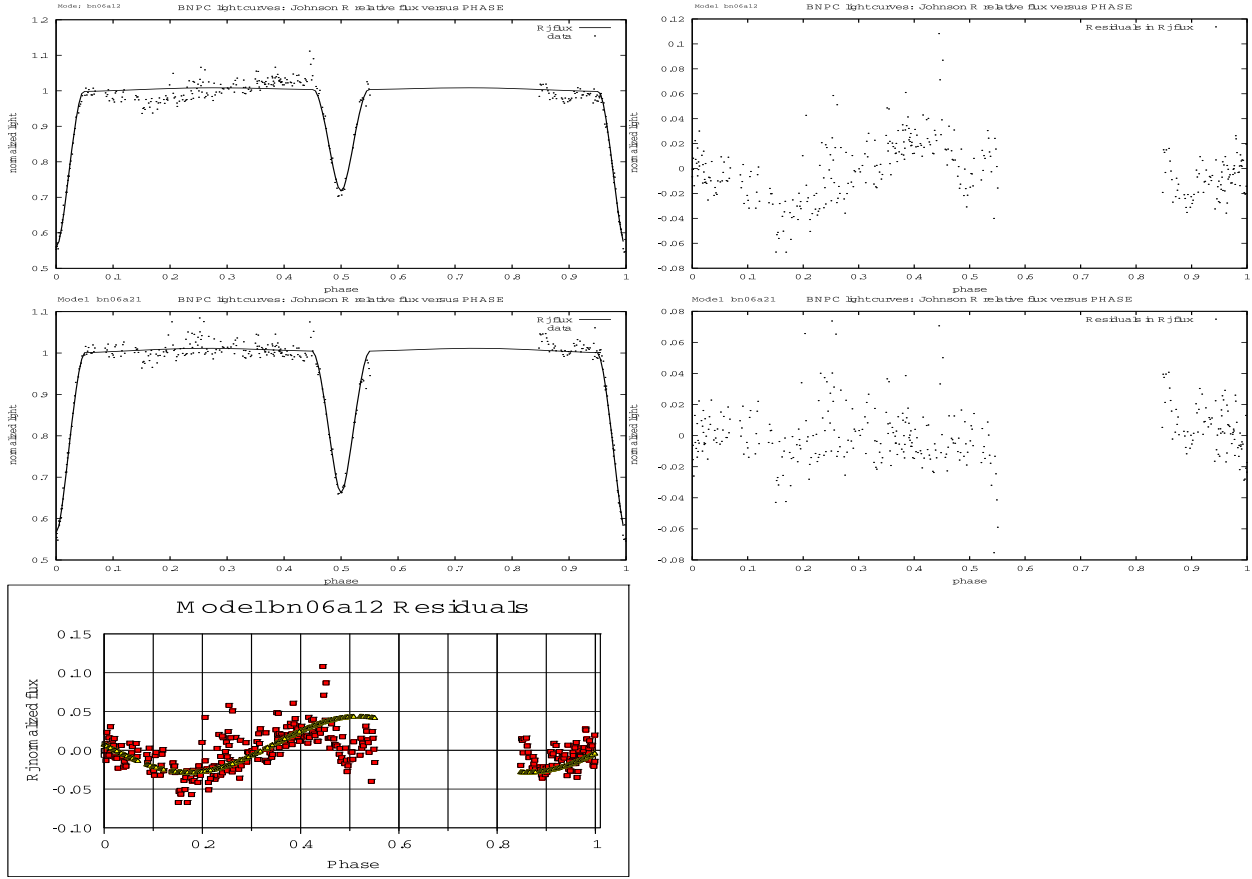}
\caption{
Light curve residuals for RAO1-06, model bn06a13 (with residuals included), top;  for RAO1-06s,
model bn06a21 (with a sinusoid fitting to the residuals removed), middle; and the empirical sinusoidal 
fitting of the residuals, bottom. See text for details.
\label{solutions9}}
\end{figure}

Analyses of RAO1-22 failed to yield a converged solution in any mode until the SA program was
invoked.  Although running time in the WD package usually required less than 10 minutes, this
run required $2^h47^m$.  It too did not achieve convergence before the stopping criteria kicked in, 
but the adjustments indicated an area of parameter space not hitherto pursued, which is, indeed, the 
great value of the simulated annealing algorithm.  The $T_2$ result indicated a higher value than
had been seen before and suggested a different course of action. The the next run, the initial epoch 
was decreased by $P/2$, so that the minima became reversed; after a number of simplex trials, a
converged solution was found for RAO1-22 with the properties shown in Table 5.  Although many
of the parameters of this system and RAO-28 are significantly different, they both required that
$T_2$ be greater than $T_1$ (at least initially for RAO1-22).  RAO1-22 exhibits a shallower and  
noisier light curve than RAO1-28 but these properties are insufficient to decide the issue.  
Thus despite the diffeences in modeled parameters, the jury is still out on separate identity 
for these variable stars.  The fittings for models bn22a20 and bn28a14 can be seen in 
Figs.~\ref{solutions5} and \ref{solutions7}, respectively.

RAO1-23 is the sole system which converged in (and thus far only in) mode 4. Its large $\Omega_2$, $T_2$, and
$L_1$ parameters indicate a hot subdwarf secondary component; the more luminous primary component
fills it Roche lobe.  The fill--out parameters were $f_1$ = +0.08, and $f_2$ = -0.003, very close to 
a mode 6 condition.  The mass ratio q = 0.92, indicates that the smaller component is slightly less 
massive.  The sparse secondary minimum is modeled as flat and somewhat deeper than the primary minimum.  
Should more data support this model, the light curve could be rephased to zero at secondary minimum, 
and the epoch altered by P/2; these changes would turn the system into a mode 5 model.  However the 
high scatter in the light curve and lack of much data in the secondary minima result in large uncertainties, 
especially in $\Omega_2$, and thus in the radius of the smaller component.  If the model survives
additional observational and analytical scrutiny, this system would be a candidate for
a multi--wavelength campaign.  The fitting for model bn23a17 can be seen in Fig.~\ref{solutions5}.

RAO1-25 was successfully modeled as an overcontact system even though much of maximum 2 is absent;
the fitting can be seen in Fig.~\ref{solutions6}.

In two cases (RAO1-30, and -33), both components appeared to be close to their
inner lobes, and after trials with modes 2, 3, and 4 or 5, runs with mode 6 were attempted.  Convergence
was obtained in this mode alone for both systems.  The result is interesting because very few systems are 
thought to be in a marginal contact state, with essentially zero thickness of the neck between them.  
If this result withstands analyses with more observations, these systems may be at cusps
in their binary evolution, and will be important to monitor in the future.  The fittings for these
systems can be seen in Figures~\ref{solutions7} and \ref{solutions8}, respectively.

\subsection{Non--solutions}

The systems for which solutions could not be obtained are: RAO1-01, -02, -03, -05, -08, -13, -15, 
-18, -34.  The light curves of RAO1-01, -03, and -05 probably lack two minima, and that of -02 as 
well as -04 show significant extra--eclipse variability, perhaps due to the proximity of known 
nearby variables.  RAO1-02 and -04 should be observed in future to disentangle the types and 
perhaps modes of variability.  As noted above, the RAO1-04 light curve distortion was not so great 
that a converged solution could not be obtained. 

In RAO1-08, a spot on the secondary star was used to model the observed difference between the maxima.
However, the insertion of a spot did not result in a convergent solution in this case.  Asymmetries in 
several other light curves (RAO1-07, -08, -10, -14, -16, -24, and -33), 
suggest the presence of starspots also, but in most of these cases the systematic difference between 
the maxima [long ago designated the ``O'Connell effect'' (Davidge \& Milone 1984\nocite{DavMil84})] 
was small enough that specific spot modeling was not needed to achieve convergence.  In some cases, 
as in RAO1-14, the asymmetry shows up mainly with a larger scatter in one maxima than in the other.

RAO1-13 appears to be a semi-detached system; the adjustment for the passband luminosity alone could not 
be brought lower than the mse in that parameter, and thus prevented convergence.

The peaked light curve of RAO1-15 could not be fitted in modes 2 or 3, and its relatively shallow light
amplitude may be due to intrinsic variability, even though previously classified as EW and EC/?. 

RAO1-18 seems to require an eccentric orbit to explain its displaced secondary minimum, but despite considerable
effort to solve it, convergence could not be obtained for this system.  In any event, the light curve is
incomplete and analysis requires more data.

The data available for RAO1-34 is rather sparse, so it is probably not surprising that convergence was
not achieved for this system.

\section{Conclusion}

To carry forward our discussion of morphological types of binaries in Section \ref{subsectypes},
we note that the proportions of converged systems in the modes 2: 4+5: 3+6 is 10:3:11 or roughly 42:13:46\%,
based not on light curve shape, but on physical morphology.  These can be compared to the ratio of these
systems from the Koch et al. (1970)\nocite{Koch70} \textit{Catalogue of Graded Photometric Studies of Close 
Binary Systems}: 30:23:47\%, proportions investigated some time ago by Linnaluoto \& Vilhu (1973)
\nocite{LinnVilhu73}.  By this measure, the detached systems appear slightly overabundant,
and the semi-detached systems slightly underabundant in our sample.  However, both samples are small ones:
the movement of one star out of one category and into another in our sample produces a relative change of 6\%.
For our sample, therefore, it is clear that the apparent difference in frequency distribution among
morphological eclipsing variable types is not significant.

In Section \ref{secdetect}, we noted the detection of likely duplicate variability in several of the
identified stars: RAO1-10 and -11; -12 and -13; -16 and -18, and -22 and -23.  Converged solutions were found 
for only one of each pair except for RAO1-10 and -11, and -22 and 28; for the former pair, the parameters agree 
within $\sim{2\sigma}$ for the most part. For the latter pair, there appear to be significant differences, 
casting some doubt on the duplicate hypothesis for this pair.
The light curve of RAO1-10 was also found to converge when run in mode 3, but with larger fitting error.  We
therefore report only the detached solution elements for this system in Table 5.

The low angular resolution of the BNPC frames compels us to treat the images
as crowded fields with single-aperture image analysis.  Therefore, the
preferred treatment for precise image analysis, multi-aperture photometry, that 
involves a ``curve-of-growth'' procedure (Stetson 1987)\nocite{Ste98} that involves
an extrapolation to include all of a star's light, cannot be applied except for 
individual cases.  
A more generally applicable approach is point-spread-function (\textit{PSF}) 
fitting photometry, a method 
that needs to be applied in order to remove the effect of neighboring stars.
The clear need for this is illustrated by five of our discoveries.
The difficulty in implementing application of this process generally is that 
the PSF is variable not only across the field but also from frame to frame, 
possibly as a result of instrumental flexure as the BNPC and the detector change 
hour angle.  However, seeing variation contributes heavily to this large
point spread function variation and this suggests that moving the instrument 
to a site with systematically better seeing could improve the central FWHM, 
and decrease some of its variation, by up to 50\%.

In later image frames than we have discussed here, we have taken shorter 
exposures as well as longer exposures, to improve the dynamical range of 
detections.  These later exposures will permit us to find
longer-period variables.  As with other searches, our period search method
precludes eruptive or irregular variable stars.  Many of these objects may 
show up, however, in a longer time base of observations.  A larger database
will also be able to test the hypothesis that our search methods appear
to favor shorter--period systems.

Despite formidable difficulties, we have detected in our first run in 
this first field 35 variable stars, of which 25 appear to be new discoveries.
In later reports, we will discuss other fields and results of analyses of 
data with longer time-bases.

Follow-up multi-color photometry is planned for all the variables with the
newly refurbished 0.4-m and 1.8-m telescopes of the RAO, and radial velocity
spectroscopy elsewhere.  These will permit more complete light curve analysis 
to be carried out for all classes of variables found here in order to determine 
their fundamental properties, and to confirm the interesting discoveries about some 
of them.

To aid in those further studies, we have performed preliminary analyses of the 
eclipsing systems for which both minima are observed.  These may be useful as starting 
values for more thorough analyses.  Although the results must be regarded as preliminary, 
the analyses have identified a number of potentially interesting 
targets for further study.  These include the apparent double--contact systems RAO1-30
and -33, the detached system RAO1-06 with sinusoidal light curve variation, 
and the apparent subdwarf semi--detached system RAO1-23.

For five variables that were previously reported, namely RAO-06, -07, -12,
-21, and -30 (see Table~\ref{knownlist} for the literature names of these objects), 
we present the first light curve analyses. In each case, we confirm and refine
the classification.

Finally, for those who desire to carry out their own analyses, the reduced data may be 
found at the website:

\url{www.ucalgary.ca/~williamd/data/BNPC\_paper1}

\acknowledgments
The BNPC was developed thanks primarily to an Alberta government Research 
Infrastructure grant on which EFM was the Principal Investigator.  The
grant was awarded under the Innovation and Science research Investments Program.
This program required substantial partner contributions, in the present
case this included a substantial contribution from university and NSERC 
grants to A. Hildebrand of the University of Calgary Geology and Geophysics Department.  
DFM Engineering, Boulder, Colorado, performed the principal 
upgrades to the instrument and local engineering support, optical and mechanical
work, and testing were carried out by Michael Mazur (the project engineer and 
consultant) and observer and data analyst Rob Cardinal.  The Faculty of Science
Workshop did the local machining.  The FLI CCD 
camera was provided under a cooperative hardware exchange agreement with Prof. 
Peter Brown of the University of Western Ontario.  Additional software improvements 
were provided by one of us (MDW), who also developed the scripts and carried out 
almost all of the observing and most of the image processing.  
Assistance in the observing, image processing, and reduction of BNPC 
data have been provided to date by undergraduate astrophysics majors and summer 
students Andrew Pon, Tyler Lenhardt, Daekwan Kim, Caleb Sundstrom, and Robert 
J. Doonan.  Operational support for this program has been provided by an NSERC 
grant to EFM and by the Department of Physics \& Astronomy, which also 
maintains the BNPC and the RAO.  
The use of IRAF software developed at the Kitt Peak National Observatory 
and the DAOPHOT package developed by Peter Stetson (1987)\nocite{Ste98} are acknowledged.
This research has made use of the SIMBAD database, operated at 
CDS, Strasbourg, France, the General Catalogue of Variable Stars, 
maintained by the Sternberg Astronomical Institute, Moscow, Russia, and the
VSX on--line variable star archive of the AAVSO.  We are grateful to an anonymous referee
for alerting us to the latter database, and for other helpful suggestions.

\clearpage

\clearpage

\clearpage

\clearpage

\clearpage

\clearpage

\clearpage

\clearpage

\clearpage

\clearpage

\clearpage

\clearpage

\clearpage

\clearpage

\clearpage

\clearpage

\clearpage

\clearpage

\clearpage

\clearpage

\clearpage

\begin{deluxetable}{rclrr}
\tabletypesize{\scriptsize}
\tablewidth{0pt}
\tablecaption{Summary of RAO BNPC Data in all RAO fields\label{all_plates}}
\tablehead{\colhead{RAO Field}&\colhead{center $\alpha$ \& $\delta$} & \colhead{JDN at 0h UT} & \colhead{Exposure (seconds)}&\colhead{Number of Frames}}
\startdata
 1 & 22:03:24 +18:54:32 & 2453256.5 & 60 & 7\\
 1 & 22:03:24 +18:54:32 & 2453271.5 & 60 & 95\\
 1 & 22:03:24 +18:54:32 & 2453272.5 & 60 & 70\\
 1 & 22:03:24 +18:54:32 & 2453281.5 & 60 & 55\\
 1 & 22:03:24 +18:54:32 & 2453282.5 & 60 & 75\\
 1 & 22:03:24 +18:54:32 & 2453283.5 & 60 & 80\\
 7 & 06:27:00 +52:30:00 & 2453305.5 & 60 & 30\\
 8 & 06:27:00 +48:30:00 & 2453305.5 & 60 & 30\\
 1 & 22:03:24 +18:54:32 & 2453305.5 & 60 & 36\\
 2 & 22:03:24 +22:54:32 & 2453305.5 & 60 & 36\\
 3 & 00:27:25 +25:36:44 & 2453310.5 & 60 & 45\\
 4 & 00:27:25 +29:36:44 & 2453310.5 & 60 & 45\\
 5 & 03:19:46 +36:53:48 & 2453313.5 & 60 & 52\\
 6 & 03:19:46 +32:53:48 & 2453313.5 & 60 & 52\\
 9 & 01:57:54 +37:39:28 & 2453965.5 & 60 & 42\\
 10 & 18:01:09 +02:54:02 & 2453965.5 & 60 & 30\\
 10 & 18:01:09 +02:54:02 & 2453966.5 & 60 & 42\\
 9 & 01:57:54 +37:39:28 & 2453972.5 & 60 & 27\\
 10 & 18:01:09 +02:54:02 & 2453972.5 & 60 & 18\\
 10 & 18:01:09 +02:54:02 & 2453973.5 & 60 & 12\\
 9 & 01:57:54 +37:39:28 & 2453974.5 & 60 & 6\\
 10 & 18:01:09 +02:54:02 & 2453974.5 & 60 & 24\\
 1 & 22:03:24 +18:54:32 & 2453981.5 & 60 & 40\\
 2 & 22:03:24 +22:54:32 & 2453981.5 & 60 & 40\\
 1 & 22:03:24 +18:54:32 & 2453982.5 & 60 & 53\\
 2 & 22:03:24 +22:54:32 & 2453982.5 & 60 & 53\\
 10 & 18:01:09 +02:54:02 & 2453981.5 & 60 & 36\\
 12 & 12:22:39 +25:49:28 & 2454226.5 & 30 & 6\\
 10 & 18:01:09 +02:54:02 & 2454270.5 & 10,60 & 4,21\\
 10 & 18:01:09 +02:54:02 & 2454288.5 & 10,60 & 5,29\\
 1 & 22:03:24 +18:54:32 & 2454307.5 & 60 & 9\\
 2 & 22:03:24 +22:54:32 & 2454307.5 & 60 & 9\\
 1 & 22:03:24 +18:54:32 & 2454308.5 & 60 & 28\\
 2 & 22:03:24 +22:54324 & 2454308.5 & 60 & 28\\
 10 & 18:01:09 +02:54:02 & 2454308.5 & 10,60 & 6,12\\
 10 & 18:01:09 +02:54:02 & 2454309.5 & 10,60 & 9,17\\
 1 & 22:03:24 +18:54:32 & 2454342.5 & 60 & 37\\
 2 & 22:03:24 +22:54:32 & 2454342.5 & 60 & 37\\
 1 & 22:03:24 +18:54:32 & 2454343.5 & 60 & 31\\
 2 & 22:03:24 +22:54:32 & 2454343.5 & 60 & 31\\
 1 & 22:03:24 +18:54:32 & 2454363.5 & 60 & 31\\
 2 & 22:03:24 +22:54:32 & 2454363.5 & 60 & 31\\
 11 & 04:25:02 +16:59:20 & 2454455.5 & 30,60,120 & 37,36,34\\
  7 & 06:27:00 +52:30:00 & 2454456.5 & 60 & 86\\
  8 & 06:27:00 +48:30:00 & 2454456.5 & 60 & 86\\
 11 & 04:25:02 +16:59:20 & 2454463.5 & 30,60,120 & 18,18,18\\
 11 & 04:25:02 +16:59:20 & 2454465.5 & 30,60,120 & 22,21,21\\
  7 & 06:27:00 +52:30:00 & 2454466.5 & 60 & 28\\
  8 & 06:27:00 +48:30:00 &  2454466.5 & 60 & 28\\
 11 & 04:25:02 +16:59:20 & 2454467.5 & 30,60,120 & 47,47,47\\
 12 & 12:22:39 +25:49:28 & 2454594.5 & 30,60,120 & 16,16,16\\
 12 & 12:22:39 +25:49:28 & 2454607.5 & 30,60,120 & 6,6,6\\

\enddata
\end{deluxetable}

\clearpage

\begin{deluxetable}{ccc}
\tabletypesize{\scriptsize}
\tablewidth{0pt}
\tablecaption{Summary of Exposures for RAO Field 1 Data Reported in this Work\label{observationslist}}
\tablehead{\colhead{JD at 0h UT}&\colhead{Length of Observing (hours)}&\colhead{Number of Observations}}
\startdata
2453271.5 & 6.1 & 92\\
2453272.5 & 4.8 & 56\\
2453281.5 & 1.7 & 24\\
2453282.5 & 5.3 & 74\\
2453283.5 & 4.9 & 78\\
2453305.5 & 1.1 & 16\\
\enddata
\end{deluxetable}

\clearpage

\begin{deluxetable}{rrrrllllrl}
\tabletypesize{\scriptsize}
\rotate
\tablewidth{0pt}
\tablecaption{Stars in RAO Field 1 in which we have detected variability.\label{starlist}}
\tablecolumns{10}
\tablehead{\colhead{Star}&\colhead{$\alpha$}&\colhead{$\delta$}&\colhead{Mag.}&\colhead{Auto Period}&\colhead{Final Period}&\colhead{Epoch}&\colhead{Depth(s)}&\colhead{rms}&\colhead{LC Type}\\
RAO1-& & & & & \colhead{(days)} & \colhead{(days)} & \colhead{(mag)} & \colhead{(mag)}}
\startdata
01 & 22 03 47.7 & 19 09 14 & 12.27 & 0.3349 & \dots         & 2453271.87$\pm${0.08}  & 0.10$\pm${0.04}, N/A        & 0.03 & Algol \\
02 & 22 01 42.6 & 17 28 44 & 12.55 & 3.4306 & 1.4493$\pm${0.0029} & 2453283.74$\pm${0.05}  & 0.26$\pm${0.03}, N/A        & 0.02 & Algol \\
03 & 22 07 36.2 & 20 20 09 & 12.33 & 0.4499 & 0.8982$\pm${0.0024} & 2453271.81$\pm${0.04}  & 0.27$\pm${0.05}, N/A        & 0.03 & Algol \\
04 & 21 53 47.9 & 18 31 26 & 11.63 & 0.3179 & 0.6155$\pm${0.0029} & 2453272.79$\pm${0.03}  & 0.36$\pm${0.02}, 0.26$\pm${0.02} & 0.02 & Algol \\
05 & 22 11 35.5 & 19 34 30 & 12.94 & 0.5899 & 1.181         & 2453271.81$\pm${0.03}  & 0.36$\pm${0.03}, N/A        & 0.02 & Algol \\
06 & 22 03 30.2 & 19 39 11 & 12.97 & 0.3852 & 1.1368957     & 2453282.68$\pm${0.01}  & 0.59$\pm${0.04}, 0.39$\pm${0.03} & 0.03 & Algol \\
07 & 22 08 03.8 & 18 37 42 & 12.56 & 0.4518 & 0.4337$\pm${0.0012} & 2453283.81$\pm${0.21} & 0.27$\pm${0.02}, 0.12$\pm${0.02} & 0.01 & $\beta$ Lyrae \\
08 & 22 00 54.6 & 20 33 08 & 13.00 & 0.2756 & 0.5510$\pm${0.0021} & 2453271.78$\pm${0.05}  & 0.27$\pm${0.03}, 0.17$\pm${0.02} & 0.02 & $\beta$ Lyrae \\
09 & 22 06 30.9 & 18 57 31 & 13.11 & 0.5563 & 0.5544$\pm${0.0019} & 2453283.73$\pm${0.04}  & 0.28$\pm${0.02}, 0.10$\pm${0.04} & 0.02 & $\beta$ Lyrae \\
10\tablenotemark{a} & 21 57 06.9 & 19 12 16 & 14.37 & 0.1688 & 0.4062$\pm${0.0018} & 2453283.76$\pm${0.03} & 0.33$\pm${0.10}, 0.24$\pm${0.07} & 0.07 & $\beta$ Lyrae \\
11\tablenotemark{a} & 21 57 05.5 & 19 11 48 & 14.02 & 0.2032 & 0.4063$\pm${0.0009} & 2453272.79$\pm${0.03} & 0.42$\pm${0.08}, 0.27$\pm${0.06} & 0.05 & $\beta$ Lyrae \\
12\tablenotemark{a} & 22 04 00.8 & 19 32 51 & 13.05 & 0.3183 & 0.4835$\pm${0.07} & 2453271.75$\pm${0.02} & 0.60$\pm${0,04}, 0.23$\pm${0.03} & 0.03 & $\beta$ Lyrae \\
13\tablenotemark{a} & 22 03 59.6 & 19 33 05 & 12.97 & 0.2468 & 0.4835$\pm${0.0007} & 2453271.75$\pm${0.01} & 0.61$\pm${0.03}, 0.21$\pm${0.02} & 0.02 & $\beta$ Lyrae \\
14 & 22 02 18.0 & 18 32 38 & 12.73 & 0.2199 & 0.4395$\pm${0.0015} & 2453271.78$\pm${0.25} & 0.04$\pm${0.02}, 0.04$\pm${0.03} & 0.01 & W UMa \\
15 & 21 55 01.6 & 20 20 22 & 11.44 & 0.1551 & 0.3736$\pm${0.0016} & 2454282.82$\pm${0.17} & 0.06$\pm${0.03}, 0.06$\pm${0.02} & 0.02 & W UMa \\
16\tablenotemark{a} & 21 59 08.2 & 17 44 16 & 12.52 & 0.1674 & 0.3299$\pm${0.0013} & 2453283.76$\pm${0.12} & 0.06$\pm${0.04}, 0.06$\pm${0.04} & 0.03 & W UMa \\
17 & 21 59 13.3 & 18 22 01 & 12.20 & 0.8762 & 0.5708$\pm${0.0020} & 2453272.73$\pm${0.82} & 0.11$\pm${0.01}, 0.10$\pm${0.01} & 0.008 & W UMa \\
18\tablenotemark{a} & 21 59 05.5 & 17 44 30 & 12.58 & 0.2463 & 0.4928$\pm${0.0021} & 2453283.75$\pm${0.05}  & 0.15$\pm${0.06}, 0.16$\pm${0.04} & 0.04 & W UMa \\
19 & 22 01 06.2 & 19 58 21 & 14.00 & 0.1718 & 0.3437$\pm${0.0010} & 2453283.74$\pm${0.06}  & 0.15$\pm${0.06}, 0.12$\pm${0.04} & 0.04 & W UMa \\
20 & 22 00 14.3 & 18 57 23 & 13.73 & 0.1395 & 0.3244$\pm${0.0009}  & 2453271.88$\pm${0.03}  & 0.16$\pm${0.11}, 0.13$\pm${0.08}& 0.07 & W UMa \\
21 & 22 05 41.9 & 19 55 08 & 11.41 & 0.1518 & 0.3037$\pm${0.0006}  & 2453272.87$\pm${0.08}  & 0.19$\pm${0.01}, 0.17$\pm${0.01} & 0.009 & W UMa \\
22\tablenotemark{a} & 21 57 58.6 & 18 58 10 & 14.00 & 0.1816 & 0.3631$\pm${0.0010} & 2453283.69$\pm${0.09}  & 0.24$\pm${0.09}, 0.27$\pm${0.11} & 0.06 & W UMa \\
23 & 22 01 04.0 & 19 18 40 & 14.92 & 0.1752 & 0.3511$\pm${0.0015} & 2453282.72$\pm${0.03}  & 0.25$\pm${0.14}, 0.28$\pm${0.19} & 0.10 & W UMa \\
24 & 22 07 27.7 & 18 09 35 & 14.45 & 0.1305 & 0.3044$\pm${0.0011} & 2453271.89$\pm${0.05}  & 0.27$\pm${0.11}, 0.21$\pm${0.09} & 0.08 & W UMa \\
25 & 21 57 49.0 & 20 48 59 & 14.22 & 0.1704 & 0.3406$\pm${0.0015} & 2453271.74$\pm${0.04}  & 0.29$\pm${0.08}, 0.23$\pm${0.09}  & 0.06 & W UMa \\
26 & 22 08 04.5 & 18 40 13 & 13.86 & 0.1898 & 0.3797$\pm${0.0009}  & 2453271.88$\pm${0.03}  & 0.31$\pm${0.06}, 0.25$\pm${0.05}  & 0.04 & W UMa \\
27 & 22 09 19.1 & 20 22 27 & 14.06 & 0.1580 & 0.3160$\pm${0.0012} & 2453282.71$\pm${0.03}  & 0.31$\pm${0.08}, 0.23$\pm${0.07}  & 0.05 & W UMa \\
28\tablenotemark{a} & 21 57 56.0 & 18 58 04 & 13.37 & 0.1816 & 0.3631$\pm${0.0005} & 2453272.79$\pm${0.03} & 0.35$\pm${0.03}, 0.35$\pm${0.04} & 0.02 & W UMa \\
29 & 22 05 24.3 & 20 41 47 & 14.72 & 0.1152 & 0.2307$\pm${0.0007}  & 2455272.74$\pm${0.01}  & 0.38$\pm${0.12}, 0.30$\pm${0.10} & 0.08 & W UMa \\
30 & 21 54 29.8 & 19 03 48 & 12.75 & 0.2955 & 0.5906$\pm${0.0013} & 2453283.77$\pm${0.04}  & 0.39$\pm${0.04}, 0.37$\pm${0.03} & 0.03 & W UMa \\
31 & 22 07 22.6 & 19 38 47 & 14.31 & 0.1448 & 0.3390$\pm${0.0012} & 2453282.85$\pm${0.25} & 0.40$\pm${0.10}, 0.26$\pm${0.09} & 0.06 & W UMa \\
32 & 22 03 43.0 & 19 53 10 & 14.73 & 0.1322 & 0.2644$\pm${0.0007}  & 2453271.73$\pm${0.01}  & 0.59$\pm${0.19}, 0.41$\pm${0.14} & 0.12 & W UMa \\
33 & 22 00 22.2 & 18 47 00 & 14.75 & 0.1805 & 0.3610$\pm${0.0011} & 2453271.86$\pm${0.01}  & 0.63$\pm${0.19}, 0.50$\pm${0.19} & 0.13 & W UMa \\
34 & 22 01 49.4 & 17 59 41 & 11.45 & 0.2619 & 0.3439$\pm${0.0018} & 2453271.71$\pm${0.18} & 0.24$\pm${0.03},  0.22$\pm${0.03} & 0.09 & W UMa \\
35 & 22 03 28.2 & 20 24 44 & 13.70 & 1.2664 & 0.5876$\pm${0.0029} & 2453283.80$\pm${0.09}  & 0.30$\pm${0.04}  & 0.03 & RR Lyrae \\
36 & 22 05 29.8 & 17 21 33 & 13.08 & 0.6799 & 0.6790$\pm${0.0028} & 2453272.78$\pm${0.19} & 0.37$\pm${0.03}  & 0.02 & RR Lyrae \\
37 & 22 03 04.0 & 16 44 18 & 13.32 & 0.6527 & 0.6212$\pm${0.0018} & 2453271.83$\pm${0.07}  & 0.71$\pm${0.07}  & 0.05 & RR Lyrae \\
38 & 22 01 23.7 & 18 24 49 & 14.63 & 0.5243 & 0.5251$\pm${0.0015} & 2453271.76$\pm${0.03}  & 0.83$\pm${0.13} & 0.09 & RR Lyrae \\
39 & 22 09 19.0 & 19 43 56 & 13.49 & 0.4349 & 0.4346$\pm${0.0009}  & 2453283.80$\pm${0.04}  & 0.90$\pm${0.03}  & 0.02 & RR Lyrae \\ 
\enddata
\tablenotetext{a}{Star pairs 10 \& 11, 12 \& 13, 16 \& 18, 22 \& 28 are close neighbors on 
the BNPC frames. In each case, it is likely that only one of each pair is sensibly variable 
and that the variability in the other is due to light contamination}.
\end{deluxetable}

\clearpage

\begin{deluxetable}{lllcccccc}
\tabletypesize{\scriptsize}
\rotate
\tablewidth{0pt}
\tablecaption{Known variable stars in RAO Field 1.\tablenotemark{a}
\label{knownlist}}
\tablehead{\colhead{Star}&\colhead{$\alpha$}&\colhead{$\delta$}
&\colhead{Lit. type\tablenotemark{b}}&\colhead{$\lambda$\tablenotemark{c} Mag.}&\colhead{Lit. Period}
&\colhead{Ref}&\colhead{Status\tablenotemark{d}}&\colhead{RAO1-}}
\startdata
GSC 01683-01853 	   & 21 53 23.67 & +17 30 20.2 & U    & V 13.00-? & 5.276 & 5 & DP & \dots \\ 
 = 1SWASP J21532.65+173020.3\\
TYC 1683-00877-1\tablenotemark{e} & 21 53 45.20 & +18 31 58.5 & EW      & w 11.40 & 0.587 & 2 & N  & \dots \\ 
USNO-B1.0 1080-00687717    & 21 53 58.91 & +18 02 23.9 & DSCT & w 12.39-? & 0.122 & 2 & DN & \dots \\
USNO-B1.0 1090-00579466    & 21 54 29.83 & +19 03 52.0 & EW?  & w 12.88-? & 0.566 & 2 & DY & 30 \\
 = ASAS 215430+1903.8?	   & 21 54 30    & +19 03 49   & EC   & V 12.76-13.26 & 0.591 & 6 & (DY) & (30) \\
TYC 1687-1479-1 	   & 21 55 01.25 & +20 20 25.7 & EW   & w 12.05-? & 0.342 & 2 & DY & 15 \\
 = ASAS 215501+2020.5? 	   & 21 55 01.00 & +20 20 30   & EC/? & V 11.51-11.79 & 0.274 & 10 & (DY) & (15) \\ 
USNO-B1.0 1067-00619020    & 21 58 16.32 & +16 45 58.4 & U    & w 12.72-? & 0.318 & 2 & DP & \dots \\
NSVS 2158307+163525	   & 21 58 30.14 & +16 35 25.4 & M    & V 9.30-11.97  & 282.  & 7 & DY?  & \dots \\
USNO-B1.0 1067-0619189     & 21 58 54.40 & +16 43 37.4 & U    & w 12.50-? & 0.443 & 2 & DY & \dots \\
USNO-B1.0 1077-00719427    & 21 59 05.41 & +17 44 32.6 & EW   & w 12.91-? & 0.407 & 2 & DY & 18 \\
TSVSC1 N-N002201313-80-67-2& 21 59 44.84 & +19 30 25.5 & U    & V 14.88-? & 0.698 & 8 & DP & \dots \\
TSVSC1 N-N002201123-57-67-2& 22 01 23.65 & +18 24 57.8 & U    & V 14.44-? & 0.525 & 8 & DY & 38 \\
USNO-B1.0 1074-00692109\tablenotemark{f} & 22 01 39.78 & +17 28 33.7 & U& w 12.54-? & 0.407 & 2 & DP & \dots \\
TYC 1684-522-1 		   & 22 01 49.27 & +17 59 42.1 & EW   & w 12.12-? & 0.415 & 2 & DY & 34 \\
 = GSC 01624-00522	   & 22 01 49.25 & +17 59 42.0 & EW   & V 11.77-12.08 & 0.415 & 11 & (DY) & (34) \\
 = ASAS 220149+1759.7 \\
USNO-B1.0 1079-00712902    & 22 01 51.22 & +17 58 53.7 & U    & w 12.56-?  & 0.207 & 2 & DP & \dots \\
TYC 1684-23-1 		   & 22 01 54.39 & +18 10 31.4 & E    & w 12.61-?  & 0.830 & 2 & DP & \dots \\ 
ASAS 220304+1644.4	   & 22 03 04.   & +16 44 24   & RRAB & V 13.07-14.11 & 0.621 & 6 & DY & 37 \\
USNO-B1.0 1096-00579000    & 22 03 30.28 & +19 39 12.8 & EA   & w 13.76-?  & 1.137 & 2 & DY & 6  \\
 = NSVS 11747875 \\
NSVS 11748364		   & 22 03 59.56 & +19 33 07.2 & EA/EB& w 13.46-? & 0.484 & 7 & DY & 12 \\
ASAS 220539+1721.1	   & 22 05 39.   & +17 21 06.  & ED   & V 12.40-12.94 & 2.805 & 6& DP & \dots \\
USNO-B1.0 1099-00576195    & 22 05 41.83 & +19 55 10.9 & EW   & w 12.13-?  & 0.310 & 2 & DY & 21 \\
 = ASAS 220542+1955.2 \\
USNO-B1.0 1085-00593094    & 22 08 25.85 & +18 34 57.4 & EW   & w 11.46-? & 0.349 & 2 & DN & \dots \\
NSVS 11753088		   & 22 08 37.97 & +18 37 44.1 & E    & V 12.89-13.40 & 0.434 & 7 & DY & 7 \\
NSVS 11753752		   & 22 09 19.10 & +19 43 57.0 & RRAB & V 13.30-14.20 & 0.770 & 7 & DY & 39 \\
\enddata
\tablenotetext{a}{ This list was compiled the help of the SIMBAD, GCVS, and SVX data bases.}
\tablenotetext{b}{ Variability types: $U$ (unspecific), $E$ (eclipsing), $EA$ (Algol), $ED$ (detached eclipsing), $EB$ ($\beta$ Lyr), 
$EW$ (W UMa), $DSCT$ ($\delta$ Sct), $RRAB$ (asymmetric light curve RR Lyrae), $LB$ (variety of long--period variable), $M$ (Mira)}
\tablenotetext{c}{ Passband for magnitude range; $w$ = white-light.} 
\tablenotetext{d}{ The detection results are: $D$ (star detected); $Y$ (variability detected); $P$ (variability detected, 
but classified as systematic noise);\\
  $N$ (no variability detected).}
\tablenotetext{e} {Near, but not identical to, RAO1-04; overexposed in most images}
\tablenotetext{f} {Near, but not identical to, RAO1-02.}
\tablerefs{(1) Henry et al. (2000)\nocite{Henry00}; (2) Kane et al. (2005)\nocite{Kane05}; (3) Kazarovets et al. 
(1999)\nocite{Kazarovets99}; (4) Henry \& Henry (2000)\nocite{HenryandHenry00}; (5) Norton et al. (2007)\nocite{Norton07}; 
(6) Pojman\'{n}ski et al. (2005)\nocite{Poj05}; (7) Northern Sky Variability Survey (Wo\'{z}niak et al. 
2004a,b\nocite{Woz04a}\nocite{Woz04b}, http://skydot.lanl.gov/nsvs/nsvs.php); 
(8) Damerdji et al. (2007)\nocite{Dametal07}; (9) Wils et al. (2006)\nocite{Wilsetal06}; (10) Pojman\'{n}ski (2002)\nocite{Poj02}; 
(11) Otero et al. (2006)\nocite{Oteroetal06}.}
\end{deluxetable}

\clearpage

\begin{deluxetable}{llcrrlllllll}
\tabletypesize{\scriptsize}
\rotate
\tablewidth{0pt}
\tablecaption{Preliminary Light Curve Solution Parameters of Detected Eclipsing Systems\tablenotemark{a}\label{solutionslist}}
\tablehead{\colhead{RAO1-} & \colhead{nos.} & \colhead{Mode\tablenotemark{b}} & \colhead{$i$} & \colhead{$\Delta{T}$} 
& \colhead{$\Omega_1$} & \colhead{$\Omega_2$} & \colhead{$q$} & \colhead{$t_0$} & \colhead{$P$} & \colhead{$L_1$} &  
 \colhead{$\sigma_{1}$(\={w})} \\
 & & & \colhead{deg} & \colhead{K} & & & & \colhead{2453200+} & \colhead{d} & \colhead{$4\pi$} & \\ } 
\startdata
04 & 312 & 2 & 75.08$\pm${0.34} & 619$\pm${49}   & 4.74$\pm${0.06} & 4.89$\pm${0.07} & 1.100$\pm${0.05}\tablenotemark{c} & 72.7822$\pm${0.0007} & 0.61645$\pm${0.00007} & 7.41$\pm${0.24}  & 0.02887 \\
06 & 340 & 2 & 85.55$\pm${0.13} &  658$\pm${25}  & 8.73$\pm${0.07} & 7.17$\pm${0.13} & 1.000\tablenotemark{d} & 82.6843$\pm${0.0003} & 1.13665$\pm${0.00007} & 6.88$\pm${0.17} & 0.02195 \\
06s\tablenotemark{e} & 340 & 2 & 86.23$\pm${0.10}&  389$\pm${20}   & 7.37$\pm${0.10} & 7.32$\pm${0.03} & 1.000\tablenotemark{d} & 82.6849$\pm${0.0002} & 1.13673$\pm${0.00004} & 7.00$\pm${0.12} & 0.01610 \\
07 & 309 & 5 & 56.84$\pm${0.59} & 1474$\pm${37}  & 4.48$\pm${0.03} & 4.26            & 1.326$\pm${0.024} & 83.8069$\pm${0.0007} & 0.43377$\pm${0.00004} & 7.99$\pm${0.12} & 0.01256 \\
09 & 340 & 2 & 65.62$\pm${0.77} & 1415$\pm${79}  & 4.27$\pm${0.07} & 4.71$\pm${0.17} & 1.123$\pm${0.032} & 83.7330$\pm${0.0007} & 0.55444$\pm${0.00007} & 9.54$\pm${0.41} & 0.01987 \\
10 & 310 & 2 & 68.21$\pm${1.24} & 793$\pm${143}  & 4.57$\pm${0.16} & 5.05$\pm${0.20} & 1.366$\pm${0.056} & 83.7596$\pm${0.0016} & 0.40618$\pm${0.00008} & 7.31$\pm${0.84} & 0.05655 \\
11 & 334 & 2 & 68.40$\pm${0.29} &  581$\pm${70}  & 4.63$\pm${0.25} & 4.15$\pm${0.07} & 1.229$\pm${0.042} & 72.7924$\pm${0.0009} & 0.40631$\pm${0.00004} & 5.53$\pm${0.55} & 0.03756 \\
12 & 247 & 5 & 71.63$\pm${0.41} & 1246$\pm${44}  & 4.46$\pm${0.24} & 3.97            & 1.134 		 & 71.7521$\pm${0.0005} & 0.48363$\pm${0.00003} & 7.43$\pm${0.41} & 0.02132 \\
14 & 334 & 2 & 63.55$\pm${1.64} & -351$\pm${195} & 4.40$\pm${0.16} & 5.56$\pm${0.19} & 0.887$\pm${0.088} & 71.7715$\pm${0.0026} & 0.43959$\pm${0.00011} & 7.82$\pm${0.81}  & 0.01463 \\
16 & 341 & 3 & 55.53$\pm${5.10} &  331$\pm${116} & 4.66$\pm${0.21} & 4.66            & 1.091$\pm${0.116} & 83.7531$\pm${0.0016} & 0.32986$\pm${0.00007} & 6.36$\pm${0.50}& 0.01642 \\
17 & 321 & 3 & 52.49$\pm${0.96} & 14$\pm${78}    & 3.69$\pm${0.02} & 3.69 & 0.814$\pm${0.012} & 72.7274$\pm${0.0010} & 0.57080$\pm${0.00007} & 6.49$\pm${0.14} & 0.00828 \\
19 & 340 & 2 & 55.67$\pm${1.95} &  278$\pm${181} & 4.04$\pm${0.08} & 4.87$\pm${0.16} & 1.183$\pm${0.056} & 83.7431$\pm${0.0012} & 0.34378$\pm${0.00005} & 7.31$\pm${0.59} & 0.02928 \\
20 & 339 & 2 & 63.16$\pm${0.18} &  249$\pm${153} & 4.59$\pm${0.67} & 4.64$\pm${0.68} & 1.227$\pm${0.048} & 71.8788$\pm${0.0016} & 0.32473$\pm${0.00006} & 5.86$\pm${2.56} & 0.04617 \\
21 & 280 & 3 & 62.15$\pm${0.34} &  344$\pm${29}  & 4.08$\pm${0.02} & 4.08            & 1.011$\pm${0.013} & 72.8722$\pm${0.0003} & 0.30366$\pm${0.00001} & 6.59$\pm${0.08} & 0.00915 \\
22 & 328 & 3 & 58.75$\pm${1.84} & 1465$\pm${249} & 4.21$\pm${0.06} & 4.21	     & 1.206$\pm${0.036} & 72.7814$\pm${0.0015} & 0.30274$\pm${0.00005} & 8.55$\pm${0.36} & 0.05560 \\
23 & 341 & 4 & 74.80$\pm${4.03} &-2860$\pm${3639}& 3.62 	   &13.72$\pm${8.93} & 0.918$\pm${0.240} & 82.7178$\pm${0.0017} & 0.35121$\pm${0.00169} &10.93$\pm${0.50} & 0.07914 \\
24 & 330 & 3 & 59.93$\pm${2.50} &  761$\pm${181} & 4.63$\pm${0.08} & 4.63            & 1.439$\pm${0.055} & 71.9002$\pm${0.0019} & 0.30436$\pm${0.00006} & 6.38$\pm${0.42} & 0.06406 \\
25 & 318 & 3 & 59.60$\pm${0.65} &  580$\pm${123} & 3.84$\pm${0.03} & 3.84            & 1.054$\pm${0.022} & 71.7425$\pm${0.0013} & 0.34067$\pm${0.00005} & 6.76$\pm${0.27} & 0.04400 \\
26 & 322 & 3 & 62.39$\pm${0.78} &  353$\pm${81}  & 3.92$\pm${0.02} & 3.92            & 1.103$\pm${0.015} & 71.8786$\pm${0.0218} & 0.37964$\pm${0.00004} & 6.14$\pm${0.16} & 0.02975 \\
27 & 304 & 2 & 63.59$\pm${1.08} &  567$\pm${95}  & 3.99$\pm${0.11} & 4.04$\pm${0.13} & 1.111$\pm${0.028} & 82.7054$\pm${0.0009} & 0.31590$\pm${0.00004} & 6.79$\pm${0.59} & 0.03744 \\
28 & 341 & 3 & 66.21$\pm${0.07} & -180$\pm${61}  & 4.03$\pm${0.17} & 4.03            & 1.177$\pm${0.009} & 72.7851$\pm${0.0004} & 0.36387$\pm${0.00002} & 5.06$\pm${0.10} & 0.01988 \\
29 & 310 & 2 & 65.85$\pm${0.23} & 1300$\pm${100} & 4.01$\pm${0.18} & 3.99$\pm${0.17} & 1.094$\pm${0.044} & 72.7452$\pm${0.0009} & 0.23065$\pm${0.00003} & 6.39$\pm${0.87} & 0.05714 \\
30 & 328 & 6 & 69.37$\pm${0.39} & -179$\pm${58}  & 4.73  	   & 4.73  	     & 1.635$\pm${0.159} & 83.7711$\pm${0.0005} & 0.59059$\pm${0.00004} & 4.27$\pm${0.24} & 0.01936 \\
31 & 339 & 3 & 68.01$\pm${1.06} &1179$\pm${105}  & 4.11$\pm${0.04} & 4.11            & 1.175$\pm${0.029} & 82.8494$\pm${0.0009} & 0.33901$\pm${0.00004} & 8.07$\pm${0.27} & 0.04101 \\
32 & 339 & 3 & 73.15$\pm${0.28} &  914$\pm${110} & 4.18$\pm${0.08} & 4.18            & 1.230$\pm${0.049} & 71.7297$\pm${0.0009} & 0.26445$\pm${0.00003} & 7.40$\pm${0.30} & 0.06964 \\
33 & 340 & 6 & 73.44$\pm${0.77} & 573$\pm${143}  & 3.84  	   & 3.84  	     & 1.057$\pm${0.164} & 71.8761$\pm${0.0014} & 0.36099$\pm${0.00005} & 7.00$\pm${0.08} & 0.07974 \\
\enddata
\tablenotetext{a}{ Fully Converged Solution parameters from the most recently updated version 
of the WD package of Kallrath et al. (1998)\nocite{Kaletal98}; Milone \& Kallrath (2008)\nocite{MilKal08}} 
\tablenotetext{b}{ Model modes: 2 = detached ($\Omega_{1,2}$ independent, both adjusted); 
3 = overcontact ($\Omega_2$ = $\Omega_1$, only $\Omega_1$ adjusted); 4 = semi--detached 
($\Omega_1$ fixed at Roche lobe for component 1, $\Omega_2$ adjusted); 5 = semi-detached 
($\Omega_2$ fixed at Roche Lobe, $\Omega_1$ adjusted); 6 = double contact ($\Omega_1$ = $\Omega_2$, both fixed at Roche lobes).} 
\tablenotetext{c}{ photometric mass ratio determined by grid method.}
\tablenotetext{d}{ Assumed and unadjusted; photometric mass ratio not determinable in this case.}
\tablenotetext{e}{The solution for RAO1-06s is that for Star RAO1-06 with a sinusoid representation of the residuals 
subtracted from the light curve; see text for details.}
\end{deluxetable}

\clearpage

\begin{thebibliography}{}
\bibitem[Alcock et al.(1995)]{Alcock95} Alcock, C., Allsman, R. A., Axelrod, T. S., Bennett, D. P., Cook, K. H., Freeman, K. C.,
Griest, K., Marshall, S. L., Perlmutter, S. L., Peterson, B. A., Pratt, M. R., Quinn, P. J., Rodgers, A. W., Stubbs, C. W., 
and Sutherland, W.  1995, \apj, 445, 133
\bibitem[Alonso(2004)]{Alonso04} Alonso, S.  2004, \apj, 613, L153
\bibitem[Bakos(2002)]{Bakos02} Bakos, G. \'{A}, L\'{a}z\'{a}r, J., Papp, I., S\'{a}ri, P., and
Green, E. M.  2002, \pasp, 114, 974
\bibitem[Bakos(2004)]{Bakos04} Bakos, G. \'{A}, Noyes, R. W., Kov\'{a}cs, G., Stanek, K. Z., 
Sasselov, D. D., and Domsa, I. 2004, \pasp, 116, 266
\bibitem[Bouguer(1729)]{Bou1729} Bouguer, P. 1729, Essai d'Optique sur la Gradation de la Lumiere (Paris: Claude Jombert)
\bibitem[Damerdji et al.(2007)]{Dametal07} Damerdji, Y., Klotz, A., and Bo\"{e}r, M. 2005, \aj, 133, 1470
\bibitem[Davies(1990)]{Davies90} Davies, S. R., 1990, \mnras, 244, 93
\bibitem[Davies(1991)]{Davies91} Davies, S. R., 1991, \mnras, 251, 64 
\bibitem[Davidge and Milone(1984)]{DavMil84} Davidge, T. J., \& Milone, E. F. 1984, \apjs, 55, 571 
\bibitem[Drilling \& Landolt(2000)]{drillinglandolt00}Drilling, J.S., Landolt, A. U. 2000, in Allen's Astrophysical Quantities, 
4th ed., ed. A. N. Cox (New York: Springer \& AIP)
\bibitem[Gilmore \& Zeilik(2000)]{GilZei00}Gilmore, G. F., and Zeilik, M. 2000,  in Allen's Astrophysical Quantities, 
4th ed., ed. A. N. Cox (New York: Springer \& AIP)
\bibitem[Hall(2000)]{Hall00} Hall, D. S. 2000, in 2000, in Allen's Astrophysical Quantities, 4th ed., ed. A. N. Cox (New York: 
Springer \& AIP)
\bibitem[Hardie(1962)]{Hardie62} Hardie, R. 1962, in Stars and Stellar Systems. II. 
Astronomical Techniques, ed. W. A.  Hiltner (Chicago: University of Chicago Press), 178. 
\bibitem[Henry \& Henry(2000)]{HenryandHenry00} Henry, G. W., and Henry, S. M., 2000, IBVS, 4826 
\bibitem[Henry et al.(2000)]{Henry00} Henry, G. W., Marcy, G. M., Butler, R. P., and Vogt, S. S., 2000, \apj, 529, L41
\bibitem[Kallrath et al.(1998)]{Kaletal98} Kallrath, J., Milone, E. F., Terrell, D., and Young, A. T. 1998, \apjs, 508, 308
\bibitem[Kane et al.(2005)]{Kane05} Kane, S. R., Lister, T. A., Cameron, A. C, Horne, K., James, David, Pollacco. D. L., 
Street, R. A., and Tsapras, Y., 2005, \mnras, 362, 117
\bibitem[Kazarovets et al.(1999)]{Kazarovets99} Kazarovets, E. V., Samus, N. N., Durlevich, O. V., Frolov, M. S., Antipin, S. V., 
Kireeva, N. N., and Pastukhova, E. N., 1999, IBVS, 4659, 1
\bibitem[Kraus et al. 2007]{Kraus07} Kraus, A. L., Craine, E. R., Giampapa, M. S., Scharlach, W. W. G., and Tucker, R. A. 2007,
\aj, 134, 1488
\bibitem[Koch et al.(1973)]{Koch70} Koch, R. H., Plavec, M., and Wood, F. B. 1970, A Catalogue of Graded Photometric Studies of\
Close Binaries, Publ. Univ. of Pennsylvania, Astronomical Series, 10.
\bibitem[Linnaluoto \& Vilhu(1973)]{LinnVilhu73} Linnaluoto, S., and Vilhu, O. 1973, \aap, 25, 481
\bibitem[Milone \& Robb(1983)]{Mil83} Milone, E. F. and Robb, R. M. 1983, \pasp,95, 666
\bibitem[Milone \& Kallrath(2008)]{MilKal08} Milone, E. F. and Kallrath, J. 2008, in Short-Period Ninary Stars: Observations, 
Analyses, and Results, eds. E. F. Milone, D. A. Leahy, and D. W. Hobill (Springer Science+Business Media B. V.), Astrophysics and 
Space Science Library 352, 191.  (New York: Springer)
\bibitem[Norton et al.(2007)]{Norton07} Norton, A. J., Wheatley, P. J., West, R. G., Haswell, C. A., Street, R. A., Collier, C. A.,
Christian, D. J., Clarkson, W. I., Enoch, B., Gallaway, M., Hellier, C., Horne, K., Irwin., J., Kane, S. R., Lister, T. A., 
Nicholas, J. P., Parley, N., Pollacco, D., Ryans, R., and Skillen, I., 2007, \aap, 467, 785 
\bibitem[Otero et al.(2006)]{Oteroetal06} Otero, S. A., Hoogeveen, G. J., and Wils, P. 2006, IBVS, No. 5674
\bibitem[Pojma\'{n}ski(2005)]{Poj02} Pojma\'{n}ski, G., 2002, Acta Astronomica, 52, 397 
\bibitem[Pojma\'{n}ski et al.(2005)]{Poj05} Pojma\'{n}ski, G., Pilecki, B., and Szczygie{\l}, D., 2005, Acta Astronomica, 55, 275 
\bibitem[Stetson(1978)]{Ste87} Stetson, P. 1987, \pasp, 99, 191
\bibitem[Stetson(1990)]{Ste90} Stetson, P. 1990, \pasp, 102, 932
\bibitem[Stetson(1998)]{Ste98} Stetson, P. 1998, User's Manual for DAOPHOT II. (Victoria: NRC/DAO) 
\bibitem[Udalski et al.(1993)]{Udal93} Udalski, A., Kaluzny, J., Szyma\'{n}sk, M., Kubiak, M., 
Krzeminski, W., Mateo, M., Preston, G. W., and Paczy\'{n}ski, B. 1993, Acta Astronomica, 43, 289
\bibitem[Udalski et al.(2002)]{Udal02} Udalski, A., Paczy\'{n}ski, B., Zebru\'{n}, K., Szyma\'{n}sk, M., Kubiak, M., 
Soszy\'{n}ski, I., Szewczyk, O., Wyrzykowski, L., and Pietrzy\'{n}ski, G. Acta Astronomica, 52, 1. 
\bibitem[Wesselink et al(1976)]{Wes76} Wesselink, A. J., Gibson, J. W., and Rose, J. 1976, \baas, 8, No. 4, 521
\bibitem[Williams(2001)]{Wil01} Williams, M. D., 2001, MSc Thesis (Calgary: University of Calgary)
\bibitem[Wils et al.(2006)]{Wilsetal06} Wils, P., Lloyd, C., and Bernhard, K. 2006, \mnras, 368, 1757 
\bibitem[Wo\'{z}niak, P. R. et al.(2004a)]{Woz04a} Wo\'{z}niak, P. R., Vestrand, W. T., Akerlof, C. W., Balsano, R., Bloch, J.,
Casperson, D., Fletcher, S., Gisler, G., Kehoe, R., Kinemuchi, K., Lee, B. C., Marshall, S., McGowan, K. E., McKay, T. A., 
Rykoff, E. S., Smith, D. A. Szymanski, J., and Wren, J., 2004, \aj, 127, 2436 
\bibitem[Wo\'{z}niak, P. R. et al.(2004b)]{Woz04b} Wo\'{z}niak, P. R., Williams, S. J., Vestrand, W. T., and Gupta, V. 2004,
\aj, 128, 2965
\end{thebibliography}
\end{document}